\documentclass[10pt]{article}
\usepackage{amsmath}
\usepackage{latexsym}
\usepackage{amssymb}
\usepackage{amsfonts}
\usepackage{amsthm}
\title{MINKOWSKI, SCHWARZSCHILD AND KERR \\
METRICS REVISITED}
\author{J.-F. Pommaret \\ CERMICS, Ecole des Ponts ParisTech, France  \\
jean-francois.pommaret@wanadoo.fr\\
 http://cermics.enpc.fr/$\sim$pommaret/home.html }
\date{  }
\begin{document}
\maketitle

\noindent
{\bf ABSTRACT }  \\

In recent papers, a few physicists studying Black Hole perturbation theory in General Relativity have tried to construct the initial part of a differential sequence based on the Kerr metric, using methods similar to the ones they already used for studying the Schwarzschild geometry. Of course, such a differential sequence is well known for the Minkowski metric and successively contains the Killing (10, order 1), Riemann (20, order 2), Bianchi (20, order 1 again) operators in the linearized framework, as a particular case of the {\it Vessiot structure equations}. In all these cases, they discovered that the {\it compatibility conditions} (CC) for the corresponding Killing operator were involving {\it a mixture of both second order and third order CC} and their idea has been to exhibit only a {\it minimal number of generating ones}. However, even if they exhibited a link between these differential sequences and the number of parameters of the Lie group preserving the background metric, they have been unable to provide an intrinsic explanation of this fact, being limited by the technical use of Weyl spinors, complex Teukolsky scalars, Killing-Yano tensors or the 2+2-formalism separating the two variables (t,r) from the two other angular variables of space-time. Using the formal theory of systems of partial differential equations and Lie pseudogroups, the purpose of this difficult computational paper is to provide new intrinsic differential homological methods involving the Spencer operator in order to revisit and solve these questions, not only in the previous cases but also in the specific case of any Lie group or Lie pseudogroup of transformations. Most of these new tools are now available as computer algebra packages.  \\

\vspace{2cm}

\noindent
{\bf KEY WORDS}  \\
General relativity, Killing operator, Riemann tensor, Weyl tensor, Bianchi identities, Lie algebroid, Differential sequence, Differential module, Homological algebra, Extension modules.

\newpage

\noindent
{\bf 1) INTRODUCTION}  \\

In many recent technical papers, a few physicists working on General relativity (GR) are trying to construct high order differential sequences while starting with the Kiling operator for a given metric (Minkowski, Schwarzschild, Kerr, ...) ([1,2],[15,16],[38]). The (technical) methods involved are ranging from Killing/Killing-Yano tensors, Penrose spinors, Teukolski scalars or compexified frames. \\

Meanwhile, {\it on one side}, they have the feeling that the construction of these sequences have surely to do with the fact that the corresponding Killing systems of infinitesimal lie equations have less linearly independent solutions than the $n(n+1)/2$ that were predicted by A. Eisenhart in $1949$ for nondegenerate metrics with constant Riemannian curvature on a space of dimension $n$ ([7] along results first found by E. Vessiot in $1903$ ([40,41]), that is $10$ when $n=4$ (space-time) with the Minkowski metric, and discover that, {\it perhaps for this reason}, they have to exhibit an unexpected mixture of generating compatibility conditions (CC) of order $2$ {\it and} $3$. However, {\it on another side}, they are clearly aware of the fact that {\it their results are far from being intrinsic} and cannot be adapted to other metrics or dimensions.\\

The purpose of this rather difficult paper, even though it is written in a rather self-contained computational way and illustrated by many explicit examples, is to solve this problem in its full intrinsic generality, using a few results ranging among the most delicate ones that can be found in the formal theory of systems of ordinary differential (OD) or partial differential (PD) equations and Lie pseudogroups introduced around $1970$ by D. C. Spencer and coworkers ([9],[10],[39]). It must also be noticed that these new methods, though they can be found in many books ([20-23],[25],[29],[33]) or papers ([24],[26],[27,28],[30-32]) of the author of this paper to which we shall often refer, have rarely been acknowledged by the computer algebra community and/or by physicists. The essential new "trick" is to deal with sections of jet-bundles and {\it not} with solutions of systems of OD or PD equations.  \\

In the Section $2$, we provide and illustrate the two crucial mathematical results needed for the applications presented in the Section 3. The first result, roughly discovered by M. Janet in $1920$ ([13]), describes the way to use a certain {\it prolongation/projection} (PP) procedure {\it absolutely needed} in order to transform any sufficiently regular system into a formally integrable system and, finally, even an involutive system, that is a situation where we know that the generating CC are described by a first order involutive sytem and the possibility to construct a {\it canonical} Janet or Spencer sequence. In this homological algebraic framework, the technique of diagram chasing is {\it absolutely needed} and we ask the reader to learn at least the so-called "snake" lemma in textbooks ([3],[5],[11],[17],[19],[36]). As for the second result, it is allowing to study {\it effectively} Lie operators, namely operators acting on vector fields and such that, if two vectors are solutions, their bracket is again a solution. It is thus also {\it absolutely needed} when dealing with the PP procedure because it is not possible to work out solutions in general and the use of computer algebra is therefore a {\it pure nonsense} for the intrinsic study of most Lie operators as we shall see in Section $3$.  \\

The author thanks Prof. Lars Andersson for many enlightening discussions during his last visit to the Albert Einstein Institute in Potsdam (october 23-27, 2017), in particular for bringing the problem of finding new generating CC to his attention when studying the specific Lie operators to be met in GR.  \\

\newpage

\noindent
{\bf 2) MATHEMATICAL TOOLS}  \\

 \noindent
{\bf A) FORMAL INTEGRABILITY}: \\

All operators and systems considered in this paper will have coefficients in a differential field $K$ with $n$ derivations ${\partial}_1,...,{\partial}_n$, for example the standard derivations when $K=\mathbb{Q}(x^1,...,x^n)$ is the differential field of rational functions. Starting with a given operator ${\cal{D}}$, a {\it direct problem} is to look for {\it generating compatibility conditions} (CC) described by an operator ${\cal{D}}_1$ in such a way that ${\cal{D}}_1\eta=0$ denotes the CC needed for solving (at least locally) the linar inhomogeneous system ${\cal{D}}\xi=\eta$. Going along this way, one may construct inductively a {\it differential sequence} ${\cal{D}},{\cal{D}}_1, {\cal{D}}_2,...$ of operators such that not only ${\cal{D}}_{i+1}\circ {\cal{D}}_i=0$ of course but also each operator ${\cal{D}}_{i+1} $ generates the CC of ${\cal{D}}_i$. Such a result has been found for the first time as a footnote by M. Janet in 1920 ([13]) who even proved that, under slightly additional conditions, the sequence stops at ${\cal{D}}_n$ which is thus formally surjective, when $n$ is the number of independent variables.  \\
The main problem is that, in general and though surprising it may look like, the word "{\it generating}" has no clear meaning at all, a result leading to refine the definition of a differential sequence in an intrinsic way. Apart from the Macaulay example that will be treated and revisited later on, our two favorite examples are the following ones that will also be revisited and are treated in a way adapted to the aim of this paper. We shall denote by $m$ the number of unknowns $(y^1,...,y^m)$ also called {\it differential indeterminates}, by $n$ the number of independent variables $(x^1,...,x^n)$ and by $q$ the order of the system/operator considered. We shall finally introduce the non-commutative ring $D=K[d]=K[d_1,...,d_n]$ of differential operators $P,Q,...$ with coefficients in $K$. \\ 

\noindent
{\bf   EXAMPLE 2A.1}: With $m=1, n=2, K=\mathbb{Q}$, we shall use {\it formal derivatives} $(d_1,d_2)$ when using differential modules or computer algebra but will set $d_i(d_jy))=d_j(d_iy)=d_{ij}y=y_{ij} $ for simplicity while using jet coordinates and notations. Let us consider the second order system:  \\
\[            Py\equiv y_{22}=u, \hspace{4mm}   Qy\equiv y_{12}+y=v \hspace{4mm} \Rightarrow \hspace{4mm} y_2= v_2 - u_1 \hspace{4mm} \Rightarrow \hspace{4mm} y=v- v_{12} + u_{11}  \]
There are many different ways to look at such a system. The first natural one is to say that {\it the only solution is } $y=0$ when $u=v=0$. The second one is to look for the CC that {\it must} be satisfyed by $u,v$ and we may adopt two possible presentations:  \\
\noindent
$\bullet$ Substitute $y$ and obtain the $2$ fourth order CC:  \\
\[  \left \{  \begin{array}{rcl}
A  &\equiv  &  v_{1222} -v_{22} - u_{1122} +u=0  \\
B &  \equiv & v_{1122} - u_{1112} -u_{11}=0
\end{array} \right.  \]
which are not differentially independent because one can easily check:  \\
\[   B_{12} - B - A_{11} = 0   \]
$\bullet$ However, we also have:  \\
\[  P\circ Q - Q\circ P=0 \Rightarrow  C \equiv Pv-Qu\equiv v_{22} - u_{12} -u=0   \]
that is a second order CC.  \\
Finally, we obtain:  \\
\[  A \equiv C_{12} - C , \hspace{3mm} B \equiv C_{11} \hspace{3mm} \Leftrightarrow \hspace{3mm} C\equiv B_{22} -A_{12}+A  \]
and we discover that the CC of ${\cal{D}}=(P,Q)$ are generated by $(A,B)$ but also by $C$ alone, though any student will hesitate between the two possibilities !.  \\
Refering to {\it differential homological algebra} as in ([25]) while indicating the order of an operator below its arrow, the same trivial differential module $M=0$ ({\it care}) defined by ${\cal{D}}$ has therefore two split resolutions: \\
\[   0 \rightarrow D \underset{2}{\rightarrow} D^2 \underset{4}{\rightarrow} D^2 \underset{2}{\rightarrow} D \rightarrow 0  \]
\[    0 \rightarrow D \underset{2}{\rightarrow} D^2  \underset{2}{\rightarrow} D \rightarrow 0   \]
which are quite different even if the two Euler-Poincar\'{e} characteristics both vanish because we have:  \\
\[   1- 2 + 2 - 1 =0, \hspace{1cm}  1 - 2 + 1=0  \]

\noindent
{\bf EXAMPLE 2A.2}: (Janet) Now with $m=1,n=3,q=2$ and $K=\mathbb{Q}(x^2)$, let us consider the second order system (See [23] or [25] for more details):  \\
\[       y_{33} - x^2 y_{11}=u  , \hspace{1cm}    y_{22}=v  \]
We let the reader use computer algebra and Gr\"obner bases in order to find out the two generating CC of respective order $3$ and $6$ and work out the following  possible resolution where $dim_K(M)=12$:  \\
\[   0  \rightarrow D  \rightarrow D^2 \underset{6}{\rightarrow} D^2 \underset{2}{\rightarrow} D  \rightarrow M  \rightarrow 0.  \]

The main point in this subsection is to use the three following highly non-trivial theorems (Compare [9,10] to [25], in particular p 364-366 for details) and just follow their proofs on the two previous examples but the next example found by Macaulay will provide all details.  \\

When $X$ is a manifold with $dim(X)=n$ and local coordinates $(x^1, ... ,x^n)$, we denote by $T$ the tangent bundle and by $T^*$ the cotangent bundle. If $E$ is a vector bundle over $X$ with projection $\pi$, local coordinates $(x,y^k)$ for $k=1,...,m$, we use to denote by $dim(E)=m$ the fiber dimension of $E$. A local or global section $f$ can be locally described by $y^k=f^k(x)$. Using multi-indices $\mu=({\mu}_1,...,{\mu}_n)$ with {\it length} $\mid \mu \mid={\mu}_1 + ... + {\mu}_n$, we shall denote by $J_q(E)$ the $q$-jet bundle of $E$ over $X$ with projection ${\pi}_q$ on $X$,  local coordinates $x^i,y^k_{\mu}$ and sections $f_q =(f^k_{\mu}(x))$ over $f$ or $j_q(f)=({\partial}_{\mu}f^k_(x))$. The Spencer operator $D: J_{q+1}(E) \rightarrow T^*\otimes J_q(E): f_{q+1} \rightarrow j_1(f_q) - f_{q+1}=({\partial}_if^k(x) - f^k_i(x), {\partial}_if^k_j(x) - f^k_{ij}(x), ...)$ (care to the notation)will allow to distinguish among these two types of sections. We denote by 
$S_qT^*$ the vector bundle of (completely) symmetric tensors and by ${\wedge}^rT^*$ the vector bundle of (completely) skewsymmetric tensors over $X$. We set:  \\

\noindent
{\bf DEFINITION 2A.3}: A {\it system} of order $q$ on $E$ is an open vector subbundle $R_q\subseteq J_q(E)$ with prolongations 
${\rho}_r(R_q)=R_{q+r}=J_r(R_q)\cap J_{q+r}(E)\subseteq J_r(J_q(E))$ and symbols ${\rho}_r(g_q)=g_{q+r}=S_{q+r}T^*\otimes E \cap  R_{q+r}\subseteq  J_{q+r}(E)$ only depending on $g_q\subseteq S_qT^*\otimes E$. For $r,s\geq 0$, we denote by $R^{(s)}_{q+r}\subseteq R_{q+r}={\pi}^{q+r+s}_{q+r}(R_{q+r+s})$ the projection of $R_{q+r+s}$ on $R_{q+r}$, which is thus defined by more equations in general. The system $R_q$ is said to be {\it formally integrable} (FI) if we have $R^{(s)}_{q+r}=R_{q+r}, \forall r,s\geq 0$, that is if all the equations of order $q+r$ can be obtained by means of only $r$ prolongations. The system $R_q$ is said to be {\it involutive} if it is FI with an involutive symbol $g_q$. We shall simply denote by 
$\Theta=\{ f\in E \mid j_q(f) \in R_q\}$ the "set" of (formal) solutions. It is finally easy to prove that the Spencer operator restricts to $D:R_{q+1} \rightarrow T^*\otimes R_q$. \\

Comparing the sequences obtained in the two previous examples, we may state:  \\

\noindent
{\bf DEFINITION 2A.4}: A differential sequence is said to be {\it formally exact} if it is exact on the jet level composition of the prolongations involved. A formally exact sequence is said to be {\it strictly exact} if all the operators/systems involved are FI. A strictly exact sequence is called {\it canonical} if all the operators/systems are involutive. The only known canonical sequences are the Janet and Spencer sequences that can be defined independently from each other.\\

\noindent
{\bf EXAMPLE 2A.5}: In the first example with $dim(E)=1, dim(F_0)=2$, we let the reader prove that $dim(R_{2+r})=4, \forall r\geq 0$. Hence, if $(A,B)$ is a section of $F_1$ while $C$ is a section of $F'_1$, the jet prolongation sequence:  \\
\[  0 \rightarrow R_6  \rightarrow J_6(E) \rightarrow J_4(F_0) \rightarrow F_1 \rightarrow 0  \]
\[ 0 \rightarrow 4  \rightarrow 28 \rightarrow 30 \rightarrow 2 \rightarrow 0  \]
is {\it not} formally exact because $4 - 28 + 30 -2=4 \neq 0$, while the corresponding long sequence:  \\
\[  0 \rightarrow R_{r+4}  \rightarrow J_{r+4}(E) \rightarrow J_{r+2}(F_0) \rightarrow J_r(F'_1) \rightarrow 0  \]
\[ 0 \rightarrow 4  \rightarrow (r+5)(r+6)/2 \rightarrow (r+3)(r+4) \rightarrow (r+1)(r+2)/2 \rightarrow 0  \]
is indeed formally exact because  $4 - \frac{r^2+11r+30)}{2}+ (r^2+7r+12) - \frac{(r^2+3r+2)}{2}=0  $ but {\it not} strictly exact because 
$R_2$ is quite far from being FI.  \\

With canonical projection ${\Phi}_0:J_q(E) \Rightarrow J_q(E)/R_q=F_0$, the case $r=0,s=1$ is descibed by the following commutative and exact diagram:  \\
\[  \begin{array}{rcccccccl}
    &  0  &  &  0  &  &  0  &  &   &   \\
     &  \downarrow &  & \downarrow &  &  \downarrow  &  &   &     \\
0 \rightarrow  &  g_{q+1}  & \rightarrow & S_{q+1}T^*\otimes E & \rightarrow &  T^*\otimes F_0 & \rightarrow &   h_1 &  \rightarrow 0 \\
&  \downarrow &  & \downarrow &  &  \downarrow  &  &  \downarrow    &     \\
0 \rightarrow &  R_{q+1} & \rightarrow  & J_{q+1}(E) & \rightarrow  &  J_1(F_0) &  \rightarrow &  Q_1  &  \rightarrow 0  \\
    &  \downarrow &  & \downarrow &  &  \downarrow  &  & \downarrow   &     \\
0 \rightarrow &  R_q & \rightarrow  & J_q(E) & \rightarrow  &  F_0 &  \rightarrow &  0  &  \\ 
   &  &  & \downarrow &  &  \downarrow  &  &   & \\
   &  &  &  0  && 0  && &
\end{array}   \]
Chasing in this diagram while applying the "{\it snake}" lemma, we obtain the useful {\it long exact connecting sequence}:  \\
 \[  0  \rightarrow  g_{q+1}  \rightarrow R_{q+1}  \rightarrow R_q  \rightarrow h_1 \rightarrow Q_1  \rightarrow 0  \]
which is thus connecting in a tricky way FI ({\it lower left}) with CC ({\it upper right}).  \\

Having in mind the fact that the {\it Riemann} ($g_1$ is not $2$-acyclics but $g_2$=0 is trivially involutive) and the {\it Weyl} (${\hat{g}}_1$ is not $2$-acyclic but ${\hat{g}}_2$ is $2$-acyclic $ \forall n\geq 4$ while ${\hat{g}}_3=0, \forall n\geq 3$) operators (linearization of the respective tensors) are second order, a key stone is:  \\

\noindent
{\bf THEOREM 2A.6}: If a FI system $R_q\subset J_q(E)$ is FI, then the corresponding operator ${\cal{D}}:E \stackrel{j_q}{\longrightarrow} J_q(E) \stackrel{{\Phi}_0}{\longrightarrow} J_q(E)/R_q=F_0$ admits (minimal) generating CC of order $s+1$ if $s$ is the smallest integer such that 
$g_{q+s}$ becomes $2$-acyclic.  \\

\noindent
{\bf THEOREM 2A.7}: ({\it prolongation/projection} (PP) {\it procedure}) If an arbitrary system $R_q\subseteq J_q(E)$ is given, one can {\it effectively} find two integers $r,s \geq 0$ such that the system $R^{(s)}_{q+r}$ is formally integrable or even involutive. \\

\noindent
{\bf THEOREM 2A.8}:  Starting with {\it any} operator ${\cal{D}}=E \rightarrow F_0$ of order $q$ defined over a differential field $K$ by a system $R_q\subseteq J_q(E)$, one can construct a formally exact commutative diagram:  \\
\[    \begin{array}{rccccccl}
0   \rightarrow &\Theta &\rightarrow &E &\stackrel{{\cal{D}}}{\longrightarrow } &F_0& \stackrel{{\cal{D}}_1}{\longrightarrow} &F_1    \\
       &  \parallel  & & \parallel & & \hspace{3mm} \downarrow {\cal{P}} &  &   \downarrow {\cal{Q}}    \\
0   \rightarrow& \Theta &\rightarrow &E& \stackrel{{\cal{D}}'}{\longrightarrow }& F'_0 &\stackrel{{\cal{D}}'_1}{\longrightarrow} &F'_1    
\end{array}   \]
where ${\cal{P}}$ is an injective operator of order at least $r+s$ whenever $R^{(s)}_{q+r}$ is a formally integrable system with the same solutions as $R_q$ obtained by the PP procedure and the upper sequence is formally exact while the lower sequence is strictly exact with $ord({\cal{D}}'_1)=1$ when  ${\cal{D}}'$ is involutive. \\

Starting with an arbitrary system $R_q\subset J_q(E)$, the main purpose of the next crucial example is to prove that the generating CC of the operator 
${\cal{D}}={\Phi}_0 \circ j_q : E \stackrel{j_q}{\longrightarrow}J_q(E) \stackrel{{\Phi}_0}{\longrightarrow } J_q(E)/R_q=F_0$, though they are of course fully determined by the first order CC of the final involutive system $R^{(s)}_{q+r}$ produced by the up/down PP procedure, are in general of order $r+s+1$ as we shall see in the applications but may be of strictly lower order. All diagrams and chases will be described in actual practice. We invite the reader to study similarly the first example where $R^{(4)}_2=0,{\cal{D}}'=j_2$ and ${\cal{D}}'_1=D_1$ is the first Spencer operator.\\

\noindent
{\bf EXAMPLE 2A.9} : ([18], p 40 with the first feeling of Formal Integrability) . With $m=1, n=3, q=2$, let us consider the second order linear system $R_2 \subset J_2(E)$ with $dim(E)=1, dim (R_2)= 8$ and ${par}_2=\{ y, y_1, y_2, y_3, y_{11}, y_{12}, y_{22}, y_{23}\}$ if we use jet coordinates, defined by the two inhomogeneous PD equations:  \\
\[  Py\equiv y_{33} = u, \hspace{2cm} Qy\equiv y_{13} + y_2=v  \]
As we already said, the only existing intrinsic procedure has two steps:  \\

\noindent
$\bullet$ {\it Step} 1:  First of all we have to look for the symbol $g_2$ defined by the two linear equations $v_{33}=0, v_{13}=0$. The coordinate system is not $\delta$-regular and exchanging $x^1$ with $x^2$, we get the Janet board:  \\
\[ \left \{ \begin{array}{rcl}
  v_{33} &  =  & 0  \\
  v_{23}  &  =  &  0 
  \end{array} \right. \fbox{ $\begin{array}{lll}
  1 & 2 & 3  \\
  1 & 2 & \bullet
  \end{array} $ }   \]
 It follows that $g_2$ is involutive, thus $2$-acyclic and we obtain from the main theorem ${\rho}_r(R^{(1)}_2)= R^{(1)}_{2+r}$. However, $R^{(1)}_2 \subset R_2$ with a strict inclusion because $R^{(1)}_2$ with $dim(R^{(1)}_2)= 7$ is now defined by the 3 equations:  \\
 \[    y_{33}= u , \hspace{1cm}  y_{23}=  v_3 - u_1, \hspace{1cm}  y_{13} + y_2= v   \]
We may start again with $R^{(1)}_2$ and study its symbol $g^{(1)}_2 $ defined by the 3 linear equations with Janet board:  \\
\[  \left \{ \begin{array}{lcl}
  v_{33} &  =  & 0  \\
  v_{23}  &  =  &  0 \\
  v_{13}  &  =  &  0
  \end{array} \right. \fbox{ $\begin{array}{lll}
  1 & 2 & 3  \\
  1 & 2 & \bullet  \\
  1  &  \bullet &  \bullet 
  \end{array} $ }   \]
It follows that $g^{(1)}_2$ is again involutive but we have a strict inclusion $R^{(2)}_2 \subset R^{(1)}_2$ because $dim(R^{(2}_2)=6$ as $R^{(2)}_2 $ is defined by the 4 equations with Janet board:   \\
\[  \left \{ \begin{array}{lcl}
  y_{33} &  =  & u \\
  y_{23}  &  =  &  v_3 - u_1   \\
  y_{22}  &  =  & v_2 -v_{13} + u_{11} \\
  y_{13} + y_2  &  =  &  v
  \end{array} \right. \fbox{ $\begin{array}{lll}
  1 & 2 & 3  \\
  1 & 2 & \bullet  \\
  1 & 2 & \bullet   \\
  1  &  \bullet &  \bullet 
  \end{array} $ }   \]
As $R^{(2)}_2$ is easily seen to be involutive, we achieve the PP procedure with the strict intrinsic inclusions and corresponding fiber dimensions:  \\
\[ R^{(2)}_2 \subset  R^{(1)}_2 \subset  R_2    \hspace{1cm}  \Leftrightarrow  \hspace{1cm}   6  <  7  <  8     \]
Finally, we have ${\rho}_r(R^{(2)}_2)= {\rho}_r((R^{(1)}_2)^{(1)})=({\rho}_r(R^{(1)}_2))^{(1)}=(R^{(1)}_{2+r})^{(1)}=R^{(2)}_{2+r}$.  \\

\noindent
$\bullet$ {\it Step} 2: It remains to find out the CC for $(u,v)$ in the initial inhomogeneous system. As we have used two prolongations in order to exhibit $R^{(2)}_2$, we have second order formal derivatives of $u$ and $v$ in the right members. Now, as we have an involutive system, we have first order CC for the new right members and could hope therefore for third order generating CC. However, we have successively the $4$ CC:  \\
\[  y_{233}=d_3(v_3-u_1)= d_2u \Rightarrow     v_{33}-u_{13} -u_2=0            \]
\[  y_{223}= d_3(v_2-v_{13}+u_{11})= d_2(v_3-u_1)  \Rightarrow    v_{133}  - u_{113} -u_{12}=0   \]
\[  y_{133}+y_{23}= d_3 v= d_1u +(v_3 - u_1) \Rightarrow  0=0   \]
\[  y_{123}+y_{22}= d_2v = d_1(v_3-u_1) +(v_2 -v_{13}+u_{11}) \Rightarrow   0=0        \]
It follows that we have {\it only} one second and one third order CC:  \\
\[  v_{33} -u_{13} -u_2=0    , \hspace{2cm}   v_{133} -u_{113} - u_{12}=0  \]
but, {\it surprisingly}, the {\it only} generating second order CC $v_{33} -u_{13} -u_2=0$ which is coming from the fact that the operator $P$ commutes with the operator $Q$. \\

  We finally show how FI is related to CC by means of an homological procedure known as the " {\it long exact connecting sequence} " which is the main byproduct of the so-called {\it snake lemma} used when chasing in diagrams. Needless to say that absolutely no classical procedure can produce such a result which is thus totally absent from the GR papers already quoted. First of all, let us compute the dimensions of the bundles that will be used in the following diagrams:   \ \
\[  {par}_2=\{  y, y_1, y_2, y_3, y_{11}, y_{12}, y_{22}, y_{23}  \}   \]
\[  {par}_3= \{  y, y_1, y_2, y_3, y_{11}, y_{12}, y_{22}, y_{111}, y_{112}, y_{122}, y_{222}, y_{223}\}   \]
\[ {par}_4= \{  y, y_1, y_2, y_3, y_{11}, y_{12}, y_{111}, y_{112}, y_{122}, y_{222},y_{1111}, y_{1112}, y_{1122}, y_{1222}, y_{2222}, y_{2223}\}  \]
and thus $dim(R_2)=8, \,\, dim(R_3)=12, \,\, dim(R_4)= 16$ in the following commutative and exact diagrams where $E$ is the trivial vector bundle with $dim(E)=1$ and $dim(g_{r+2})=r+4, \forall r\geq 0$:  \\
{\small  \[  \begin{array}{rcccccccl}
    &  0  &  &  0  &  &  0  &  &   &   \\
     &  \downarrow &  & \downarrow &  &  \downarrow  &  &   &     \\
0 \rightarrow  &  g_4  & \rightarrow & S_4T^*\otimes E & \rightarrow &  S_2T^*\otimes F_0 & \rightarrow &   h_2 &  \rightarrow 0 \\
&  \downarrow &  & \downarrow &  &  \downarrow  &  &  \downarrow    &     \\
0 \rightarrow &  R_4 & \rightarrow  & J_4(E) & \rightarrow  &  J_2(F_0) &  \rightarrow &  F_1  &  \rightarrow 0  \\
    &  \downarrow &  & \downarrow &  &  \downarrow  &  & \downarrow   &     \\
0 \rightarrow &  R_3 & \rightarrow  & J_3(E) & \rightarrow  &  J_1(F_0) &  \rightarrow &  0  &  \\ 
   &  &  & \downarrow &  &  \downarrow  &  &   & \\
   &  &  &  0  && 0  && &
\end{array}   \]}
{\small\[  \begin{array}{rcccccccl}
    &  0  &  &  0  &  &  0  &  &   &   \\
     &  \downarrow &  & \downarrow &  &  \downarrow  &  &   &     \\
0 \rightarrow  &  6  & \rightarrow &15 & \rightarrow & 12 & \rightarrow &  3&  \rightarrow 0 \\
&  \downarrow &  & \downarrow &  &  \downarrow  &  &  \downarrow    &     \\
0 \rightarrow &  16 & \rightarrow  & 35  & \rightarrow  & 20  &  \rightarrow &  1  &  \rightarrow 0  \\
    &  \downarrow &  & \downarrow &  &  \downarrow  &  & \downarrow   &     \\
0 \rightarrow & 12 & \rightarrow  &  20   & \rightarrow  & 8  &  \rightarrow &  0  &  \\ 
   &  &  & \downarrow &  &  \downarrow  &  &   & \\
   &  &  &  0  &  & 0  &  &  &
\end{array}   \]}
{\small \[  \begin{array}{rcccccccl}
 & 0\hspace{2mm}& & 0 \hspace{2mm} && 0\hspace{2mm}& & 0 \hspace{2mm} & \\
 & \downarrow \hspace{2mm}& & \downarrow \hspace{2mm}& & \downarrow \hspace{2mm} & & \downarrow  \hspace{2mm}&\\
0 \rightarrow & g_4 & \rightarrow & S_4T^*\otimes E & \rightarrow  & S_2T^* \otimes F_0 & \rightarrow  & h_2 & \rightarrow 0  \\
  &  \downarrow \delta  & & \downarrow \delta & & \downarrow  \delta & & \downarrow \delta  &\\
0 \rightarrow & T^*\otimes g_3 & \rightarrow & T^* \otimes S_3T^* \otimes E & \rightarrow & T^*\otimes T^*\otimes F_0 &\rightarrow & T^*\otimes h_1 & \rightarrow 0  \\
   &  \downarrow \delta  & & \downarrow \delta & & \downarrow  \delta & &\downarrow \hspace{2mm}  &   \\
0 \rightarrow & {\wedge}^2T^*\otimes g_2 & \rightarrow &{\wedge}^2T^*\otimes S_2T^* \otimes E & \rightarrow & {\wedge}^2T^*\otimes F_0 & \rightarrow  &  0 \hspace{2mm}& \\
  & \downarrow \delta &  & \downarrow \delta & & \downarrow  \hspace{2mm}& &  & \\
 0 \rightarrow  & {\wedge}^3T^*\otimes  T^* \otimes E & = &   {\wedge}^3T^*\otimes  T^* \otimes E &\rightarrow  & 0 \hspace{2mm}& & &  \\
  &\downarrow \hspace{2mm} & &\downarrow \hspace{2mm} & & & & &  \\
  & 0 \hspace{2mm} &  &  0 \hspace{2mm} &  &  &  &  &
\end{array}   \] }
{\small \[  \begin{array}{rcccccccl}
 & 0\hspace{2mm}& & 0 \hspace{2mm} && 0\hspace{2mm}& & 0 \hspace{2mm} & \\
 & \downarrow \hspace{2mm}& & \downarrow \hspace{2mm}& & \downarrow \hspace{2mm} & & \downarrow  \hspace{2mm}&\\
0 \rightarrow & 6 & \rightarrow & 15 & \rightarrow  & 12 & \rightarrow  & 3 & \rightarrow 0  \\
  &  \downarrow \delta  & & \downarrow \delta & & \downarrow  \delta & & \downarrow \delta  &\\
0 \rightarrow & 15 & \rightarrow & 30 &\rightarrow & 18 & \rightarrow& 3 & \rightarrow  0  \\
   &  \downarrow \delta  & & \downarrow \delta & & \downarrow  \delta & &\downarrow \hspace{2mm}  &   \\
0 \rightarrow & 12  & \rightarrow &18 & \rightarrow & 6 & \rightarrow  &  0 \hspace{2mm}& \\
  & \downarrow \delta &  & \downarrow \delta & & \downarrow  \hspace{2mm}& &  & \\
 0 \rightarrow  & 3 & = &   3 &\rightarrow  & 0 \hspace{2mm}& & &  \\
  &\downarrow \hspace{2mm} & &\downarrow \hspace{2mm} & & & & &  \\
  & 0 \hspace{2mm} &  &  0 \hspace{2mm} &  &  &  &  &
\end{array}   \] }

\noindent
where $S_4T^*\otimes E \simeq S_4T^*$ and $F_1\simeq Q_2$. From the snake lemma and a chase, we obtain the {\it long exact connecting sequence}:
\[  0  \rightarrow  g_4  \rightarrow R_4  \rightarrow R_3  \rightarrow h_1 \rightarrow F_1  \rightarrow 0  \]
\[  0  \rightarrow  6  \rightarrow 16 \rightarrow 12 \rightarrow 3 \rightarrow  1  \rightarrow 0  \]
relating FI ({\it lower left}) to CC ({\it upper right}). By composing the epimorphism $S_2T^* \otimes F_0 \rightarrow h_1$ with the epimorphism $h_1\rightarrow F_1$, we obtain an epimorphism $S_2T^*\otimes F_0 \rightarrow F_1$ and the long exact sequence:  \\
\[ 0 \rightarrow g_4 \rightarrow S_4T^*\otimes E \rightarrow  S_2T^*\otimes F_0 \rightarrow F_1 \rightarrow  0  \]  \\
which is nevertheless not a long ker/coker exact sequence by counting the dimensions as we have $6 - 15 + 12 -1= 2 \neq 0$.  \\
In order to convince the reader about the usefulness of these new methods, even on such an elementary example, let us prove directly the exactness of the following long exact sequence $\forall r\geq 0$:  \\
\[  0  \rightarrow R_{r+4} \rightarrow  J_{r+4}(E) \rightarrow J_{r+2}(F_0) \rightarrow  J_r(F_1) \rightarrow 0 \]
We let the reader check that $dim(R_{r+2})=4r+8, \forall r\geq 0$ and thus $dim(R_{r+4})=4r + 16, \forall r\geq 0$ as a tricky exercise of combinatorics and then use the standard formulas:  \\
\[dim(J_{r+4}(E))=(r+5)(r+6)(r+7)/6 \]
 \[dim(J_{r+2}(F_0))=(r+3)(r+4)(r+5)/3, \hspace{5mm}  dim(J_r(F_1))=(r+1)(r+2)(r+3)/6  \]
in order to check that the {\it Euler-Poincar\'e characteristics} (alternate sum of dimensions) is zero. We let the reader prove as a chasing exercise that the previous result is equivalent to prove that the following symbol sequences where $dim(g_{r+4})=r+6,\forall r\geq 0$:  \\
\[  0 \rightarrow g_{r+4} \rightarrow S_{r+4}T^* \otimes E \rightarrow \fbox{$S_{r+2}T^*\otimes F_0$} \rightarrow S_rT^*\otimes F_1 \rightarrow 0  \]
is exact everywhere but at $S_{r+2}T^*\otimes F_0$ where the cohomology has dimension $r+2$, that is:  \\ 
\[   ( 2(r+3)(r+4)/2  -  (r+1)(r+2)/2) - ((r+5)(r+6)/2 - (r+6))= dim(R_{r+3}) -(dim(R_{r+4}) - dim (g_{r+4}) )  \]
but such a method cannot be used for the more complicate examples dealing with GR that we shall find in the next section. Starting now with $R_5$ and defining $F_1$ as the cokernel of $J_5(E) \rightarrow J_3(F_0)$, we may exhibit the formally exact {\it longer} sequence (exercise):  \\
\[  0 \rightarrow \Theta \rightarrow 1 \underset 2{\rightarrow} 2 \underset 3{\rightarrow}4 \underset 1{\rightarrow} 6 \underset 1{\rightarrow} 4 \underset 1{\rightarrow} 1 \rightarrow 0  \]

Finally, refering to the general theorem, we may consider the commutative diagram:   \\
\[ \begin{array}{rcccccccl}
0 \rightarrow & \Theta & \rightarrow &  1 &  \underset{2}{\stackrel{{\cal{D}}}{\longrightarrow}} &  2  & \underset{2}{\stackrel{{\cal{D}}_1}{\longrightarrow}} &  1  &  \rightarrow 0 \\
   & \parallel &  &  \parallel &  & \hspace{3mm} \downarrow {\cal{P}} & \searrow  & \downarrow {\cal{Q}} &  \\
0 \rightarrow & \Theta & \rightarrow &  1 &  \underset{2}{\stackrel{{\cal{D}}'}{\longrightarrow}} &  4 & \underset{1}{\stackrel{{\cal{D}}'_1}{\longrightarrow}} &  4 &
\end{array}  \]
and the double prolongation diagram with $q=2,r=0,s=2, u\geq 0$ and ${\cal{D}}=\Phi \circ j_q, {\cal{D}}_1={\Psi}_1 \circ j_2$, where the two left upper downarrows are epimorphisms while the two left lower downarrows are monomorphisms: \\
\[   \begin{array}{lccccccc}
0 \rightarrow &R_{q+s+v+1} & \rightarrow &  J_{q+s+v+1}(E) & \stackrel{{\rho}_{s+v+1}(\Phi)}{\longrightarrow} &  J_{s+v+1}(F_0)  &  \stackrel{ {\rho}_{ v+1}({\Psi}_1)}{\longrightarrow}&   J_{v+1}(F_1)        \\                                                      
 & \downarrow &  &   \downarrow  &  & \downarrow &  & \downarrow  \\
 0 \rightarrow & R^{(s)}_{q+v+1}& \rightarrow & J_{q+v+1}(E)  &  \stackrel{{\rho}_{v+1}({\Phi}')}{\longrightarrow} & J_{ v+1}(F'_0)& \stackrel{{\rho}_{v}({\Psi}'_1)}{\longrightarrow} & J_v(F'_1) \\                                                                                     
   & \downarrow & &  \downarrow  &  &   \downarrow  &  &  \\
0 \rightarrow & R_{q+v+1} & \rightarrow  & J_{q+v+1}(E) & \stackrel{{\rho}_{v+1}(\Phi)}{\longrightarrow} &  J_{v+1}(F_0) &  &
\end{array}   \]
because we have been able to choose $ord({\cal{D}}_1)= r+s=s= 2$ instead of $r+s+1$ as usual and where the columns are {\it not} sequences. Chasing now in an unusual way, we may start with any $b\in B_{s+u+1}=ker({\rho}_u({\Psi}_1))$ whenever $u\geq 0$ that we can lift to $\bar{b}\in B_{s+v+1}$ because we have chosen an involutive CC system, whenever $v\geq u\geq 0$. Choosing $v=r+s+u=s+u$, we may use the fact that the central line defines a Janet sequence and is thus an exact sequence. Therefore, choosing $\bar{c}\in J_{q+v+1}(E)$ such that ${\rho}_{v+1}({\Phi}')(\bar{c})$ is the image of $\bar{b}$ in $J_{v+1}(F'_0)$, we obtain by inclusion an element $c\in J_{q+v+1}(E)=J_{q+s+u+1}(E)$ such that ${\rho}_{v+1}(\Phi)(c)$ is $b$, ending the proof. In the differential module point of view, we have the commutative and exact diagram over $D=\mathbb{Q}[d_1,d_2,d_3]$ where the upper sequence comes from a Janet sequence:  \\
\[ \begin{array}{rcccccccccl}
0 \rightarrow &D &\underset{1}{\rightarrow}  & D^4 & \underset{1}{\rightarrow} &  D^4  &  \underset{2}{\rightarrow} &  D & \rightarrow & M &   \rightarrow 0  \\
 & & &   \downarrow  &   &  \downarrow &  &   \parallel &  & \parallel &     \\
 & 0 & \rightarrow  & D & \underset{2}{\rightarrow} &  D^2 & \underset{2}{\rightarrow} &  D & \rightarrow  &M & \rightarrow 0 
\end{array}  \]
The {\it Euler-Poincar\'e characteristic}, which is equal to the {\it differential rank} $rk_D(M)$ of $M$, vanishes in both resolutions which are both called "{\it exact} " in algebraic analysis because {\it infinite jets are implicitly used}, even though the lower one is {\it not} "{\it strictly} " exact ({\it Care}, see [26] for more details).   \\

\noindent
{\bf B) ALGEBROID BRACKET}  \\

As we do not want to provide details about groupoids, we shall introduce a "copy" $Y$ ({\it target}) of $X$ ({\it source}) and define simply a Lie pseudogroup $\Gamma \subseteq aut(X)$ as a group of transformations solutions of a (in general nonlinear) system ${\cal{R}}_q$, such that, whenever $y=f(x), z=g(y) \in \Gamma$ can be composed, then $z=g\circ f (x) \in \Gamma$, $x=f^{-1}(y)\in \Gamma $ and $y=id(x)=x \in \Gamma$. Setting 
$y=x+t {\xi}(x) +...$ and passing to the limit when $t \rightarrow 0$, we may linearize the later system and obtain a (linear) system $R_q\subset J_q(T)$ such that $[\Theta,\Theta]\subset \Theta$. We may use the Frobenius theorem in order to find a generating fundamental set of {\it differential invariants} $\{{\Phi}^{\tau}(y_q)\}$ up to order $q$ which are such that ${\Phi}^{\tau}({\bar{y}}_q)={\Phi}^{\tau}(y_q)$ whenever $\bar{y}=g(y)\in \Gamma$. We obtain the {\it Lie form} ${\Phi}^{\tau}(y_q)={\Phi}_{\tau}(id_q(x))={\Phi}^{\tau}(j_q(id)(x))={\omega}^{\tau}(x)$ of ${\cal{R}}_q$.\\

Of course, in actual practice {\it one must use sections of} $R_q$ {\it instead of solutions} and we now prove why the use of the Spencer operator becomes crucial for such a purpose. Indeed, using the {\it algebraic bracket} $\{ j_{q+1}(\xi),j_{q+1}(\eta)\}=j_q([\xi,\eta]), \forall \xi,\eta\in T$, we may  obtain by bilinearity a {\it differential bracket} on $J_q(T)$ extending the bracket on $T$:\\
\[   [{\xi}_q,{\eta}_q]=\{{\xi}_{q+1},{\eta}_{q+1}\}+i(\xi)D{\eta}_{q+1}-i(\eta)D{\xi}_{q+1}, \forall {\xi}_q,{\eta}_q\in J_q(T) \]
which does not depend on the respective lifts ${\xi}_{q+1}$ and ${\eta}_{q+1}$ of ${\xi}_q$ and ${\eta}_q$ in $J_{q+1}(T)$. This bracket on sections satisfies the Jacobi identity:  \\
\[       [[{\xi}_q,{\eta}_q] ,{\zeta}_q ] + [ [ {\eta}_q,{\zeta}_q], {\xi}_q] + [Ê[ {\zeta}_q,{\xi}_q], {\eta}_q] =0    \]
and we set ([20-23],[29]):\\

\noindent
{\bf DEFINITION 2B.1}: We say that a vector subbundle $R_q\subset J_q(T)$ is a {\it system of infinitesimal Lie equations} or a {\it Lie algebroid} if $[R_q,R_q]\subset R_q$, that is to say $[{\xi}_q,{\eta}_q]\in R_q, \forall {\xi}_q,{\eta}_q\in R_q$. Such a definition can be tested by means of computer algebra. We shall also say that $R_q$ is {\it transitive} if we have the short exact sequence $0\rightarrow R^0_q \rightarrow R_q \stackrel{{\pi}^q_0}{ \rightarrow} T  \rightarrow 0$. \\

\noindent
{\bf THEOREM 2B.2}: The bracket is compatible with prolongations: \\
\[    [R_q,R_q]\subset R_q  \hspace{5mm} \Rightarrow  \hspace{5mm}  [R_{q+r},R_{q+r}]\subset R_{q+r}, \forall r\geq 0  \]

\noindent
{\bf Proof}: When $r=1$, we have ${\rho}_1(R_q)=R_{q+1}=\{ {\xi}_{q+1}\in J_{q+1}(T)\mid {\xi}_q\in R_q, D{\xi}_{q+1}\in T^*\otimes R_q\}$ and we just need to use the following formulas showing how $D$ acts on the various brackets if we set $L({\xi}_1)\zeta=[\xi,\zeta]+i(\zeta)D{\xi}_1$ (See [20] and [23] for more details):  \\
\[  i(\zeta)D\{{\xi}_{q+1},{\eta}_{q+1}\}=\{i(\zeta)D{\xi}_{q+1},{\eta}_q\}+\{{\xi}_q,i(\zeta)D{\eta}_{q+1}\} ,\hspace {4mm} \forall \zeta \in T \]  
\[ \begin{array}{rcl}
 i(\zeta)D[{\xi}_{q+1},{\eta}_{q+1}]  & =  & [i(\zeta)D{\xi}_{q+1},{\eta}_q]+[{\xi}_q,i(\zeta)D{\eta}_{q+1}]  \\
     &   &+i(L({\eta}_1)\zeta)D{\xi}_{q+1}-i(L({\xi}_1)\zeta)D{\eta}_{q+1}  
\end{array} \]
The right member of the second formula is a section of $R_q$ whenever ${\xi}_{q+1},{\eta}_{q+1}\in R_{q+1}$. The first formula may be used when $R_q$ is formally integrable. \\
\hspace*{10cm}   Q.E.D.  \\

\noindent
{\bf COROLLARY 2B.3}: The bracket is compatible with the PP procedure:    \\ 
\[   [R_q,R_q] \subset R_q \hspace{5mm}  \Rightarrow \hspace{5mm} [R^{(s)}_{q+r},R^{(s)}_{q+r} ] \subset  R^{(s)}_{q+r}, \forall r,s \geq 0   \]

\noindent
{\bf EXAMPLE 2B4}: With $n=1, q=3, X=\mathbb{R}$ and evident notations, the components of $[{\xi}_3,{\eta}_3]$ at order zero, one, two and three are defined by the totally unusual successive formulas:\\
\[    [\xi,\eta]=\xi{\partial}_x\eta-\eta{\partial}_x\xi     \]
\[    ([{\xi}_1,{\eta}_1])_x=\xi{\partial}_x{\eta}_x-\eta{\partial}_x{\xi}_x    \]
\[    ([{\xi}_2,{\eta}_2])_{xx}={\xi}_x{\eta}_{xx}-{\eta}_x{\xi}_{xx}+\xi{\partial}_x{\eta}_{xx}-\eta{\partial}_x{\xi}_{xx}   \]
\[    ([{\xi}_3,{\eta}_3])_{xxx}=2{\xi}_x{\eta}_{xxx}-2{\eta}_x{\xi}_{xxx}+\xi {\partial}_x{\eta}_{xxx}-\eta{\partial}_x{\xi}_{xxx}   \]
Affine transformations: ${\xi}_{xx}=0,{\eta}_{xx}=0\Rightarrow ([{\xi}_2,{\eta}_2])_{xx}=0 \Rightarrow [R_2,R_2]\subset R_2$.\\
Projective transformations: ${\xi}_{xxx}=0,{\eta}_{xxx}=0 \Rightarrow ([{\xi}_3,{\eta}_3])_{xxx}=0  \Rightarrow  [R_3,R_3]\subset R_3$.  \\

\noindent
{\bf EXAMPLE 2B.5}: With $n=m=2$ and $q=1$, let us consider the Lie pseudodogroup $\Gamma \subset aut(X)$ of finite transformations $y=f(x)$ such that $y^2dy^1=x^2dx^1=\alpha$. Setting $y=x+t {\xi}(x)+...$ and linearizing, we get the Lie operator ${\cal{D}}\xi={\cal{L}}(\xi)\alpha $ where ${\cal{L}}$ is the Lie derivative and the system $R_1\subset J_1(T)$ of linear infinitesimal Lie equations:\\
\[  x^2{\partial}_1{\xi}^1 + {\xi}^2=0, \hspace{1cm}  {\partial}_2{\xi}^1=0   \]
Replacing $j_1(\xi)$ by a section ${\xi}_1 \in J_1(T)$, we have:  \\
\[   {\xi}^1_1= - \frac{1}{x^2}  {\xi}^2, \hspace{1cm}  {\xi}^1_2=0   \]
Let us choose the two sections:  \\
\[  {\xi}_1= \{ {\xi}^1=0, {\xi}^2= - x^2, {\xi}^1_1=1, {\xi}^1_2=0, {\xi}^2_1=0, {\xi}^2_2=0\}\in R_1  \] 
\[    {\eta}_1= \{ {\eta}^1=x^2, {\eta}^2=0, {\eta}^1_1=0, {\eta}^1_2=- x¬2, {\eta}^2_1=0, {\eta}^2_2=1 \} \in R_1    \]
We let the reader check that $[{\xi}_1,{\eta}_1] \in R_1$. However, we have the strict inclusion $R^{(1)}_1 \subset R_1$ defined by the additional equation ${\xi}^1_1 + {\xi}^2_2=0$ and thus ${\xi}_1, {\eta}_1 \notin R^{(1)}_1$ though we have indeed $[ R^{(1)}_1, R^{(1)}_1] \subset R^{(1)}_1$, a result not evident at all because ${\xi}_1$ and ${\eta}_1$ have {\it nothing to do do} with solutions. We invite the reader to proceed in the same way with $ \beta=x^2dx^1 -x^1dx^2$ and compare.  \\

\hspace{1cm}

\noindent
{\bf C) EXTENSION MODULES}  \\

Let $D=K[d_1,...,d_n]=K[d]$ be the ring of differential operators with coefficients in a differential field $K$ of characteristic zero, that is such that $\mathbb{Q}\subset K$, with $n$ commuting derivations ${\partial}_1,...,{\partial}_n$ and commutation relations $d_ia=ad_i+{\partial}_ia,\forall a\in K$. If $y^1,...,y^m$ are $m$ differential indeterminates, we may identify $Dy^1+...+Dy^m=Dy$ with $D^m$ and consider the finitely presented left differential module $M={}_DM$ with presentation $D^p\rightarrow D^m\rightarrow M \rightarrow 0$ determined by a given linear multidimensional system with $n$ independent variables, $m$ unknowns and $p$ equations. Applying the functor $hom_D(\bullet,D)$, we get the exact sequence $0\rightarrow hom_D(M,D)\rightarrow D^m\rightarrow D^p \longrightarrow N_D \longrightarrow 0$ of {\it right differential modules} that can be transformed by a side-changing functor to an exact sequence of finitely generated {\it left differential modules}. This new presentation corresponds to the {\it formal adjoint} $ad({\cal{D}})$ of the linear differential operator $\cal{D}$ determined by the initial presentation but now with $p$ unknowns and $m$ equations, obtaining therefore a new finitely generated {\it left differential module} $N={}_DN$ and we may consider $hom_D(M,D)$ as the {\it module of equations} of the {\it compatibility conditions} (CC) of $ad({\cal{D}})$, a result not evident at first sight (See [25]). Using now a maximum free submodule $0 \longrightarrow D^l \longrightarrow hom_D(M,D)$ and repeating this standard procedure while using the well known fact that $ad(ad({\cal{D}}))={\cal{D}}$, we obtain therefore an embedding $0\rightarrow hom_D(hom_D(M,D),D)\rightarrow D^l$ of left differential modules for a certain integer $1\leq l<m$ because $K$ is a field and thus $D$ is a noetherian bimodule over itself, a result leading to $l=rk_D(hom_D(M,D))=rk_D(M)< m$ as in ([24,25]). Now, setting $t(M)=\{ m\in M \mid \exists 0\neq P\in D, Pm=0\}$, the kernel of the map $\epsilon:M\rightarrow hom_D(hom_D(M,D),D):m\rightarrow \epsilon(m)(f)= f(m),\forall f\in hom_D(M,D)$ is the {\it torsion submodule} $t(M)\subseteq M$ and $\epsilon$ is injective if and only if $M$ is torsion-free, that is $t(M)=0$. In that case, we obtain by composition an embedding $0\rightarrow M \rightarrow D^l$ of $M$ into a free module. This result is quite important for applications as it provides a (minimal) parametrization of the linear differential operator $\cal{D}$ and amounts to the controllability of a classical control system when $n=1$ ([25]). This parametrization will be called an "{\it absolute parametrization} " as it only involves arbitrary "{\it potential-like} " functions (See [4],[[25], [26-27], [31], [33],[34,35] and [43] for more details and examples, in particular that of Einstein equations).   \\

If $P=a^{\mu}d_{\mu}\in D=K[d]$, the highest value of ${\mid}\mu {\mid}$ with $a^{\mu}\neq 0$ is called the {\it order} of the {\it operator} $P$ and the ring $D$ with multiplication $(P,Q)\longrightarrow P\circ Q=PQ$ is filtred by the order $q$ of the operators. We have the {\it filtration} $0\subset K=D_0\subset D_1\subset  ... \subset D_q \subset ... \subset D_{\infty}=D$. Moreover, it is clear that $D$, as an algebra, is generated by $K=D_0$ and $T=D_1/D_0$ with $D_1=K\oplus T$ if we identify an element $\xi={\xi}^id_i\in T$ with the vector field $\xi={\xi}^i(x){\partial}_i$ of differential geometry, but with ${\xi}^i\in K$ now. It follows that $D={ }_DD_D$ is a {\it bimodule} over itself, being at the same time a left $D$-module by the composition $P \longrightarrow QP$ and a right $D$-module by the composition $P \longrightarrow PQ$. We define the {\it adjoint} map $ad:D \longrightarrow D^{op}:P=a^{\mu}d_{\mu} \longrightarrow  ad(P)=(-1)^{\mid \mu \mid}d_{\mu}a^{\mu}$ and we have $ad(ad(P))=P$. It is easy to check that $ad(PQ)=ad(Q)ad(P), \forall P,Q\in D$. Such a definition can also be extended to any matrix of operators by using the transposed matrix of adjoint operators (See [25],[27],[30],[32] for more details and applications to control theory and mathematical physics). \\

Accordingly, if $y=(y^1, ... ,y^m)$ are differential indeterminates, then $D$ acts on $y^k$ by setting $d_{\mu}y^k=y^k_{\mu}$ with $d_iy^k_{\mu}=y^k_{\mu+1_i}$ and $y^k_0=y^k$. We may therefore use the jet coordinates in a formal way as in the previous section. Therefore, if a system of OD/PD equations is written in the form:  \\
\[ {\Phi}^{\tau}\equiv a^{\tau\mu}_ky^k_{\mu}=0\] 
with coefficients $a^{\tau \mu}_k \in K$, we may introduce the free differential module $Dy=Dy^1+ ... +Dy^m\simeq D^m$ and consider the differential submodule $I=D\Phi\subset Dy$ which is usually called the {\it module of equations}, both with the {\it differential module} $M=Dy/D\Phi$ or $D$-module and we may set $M={ }_DM$ if we want to specify the ring of differential operators. Again, we may introduce the formal {\it prolongation} with respect to $d_i$ by setting:  \\
\[  d_i{\Phi}^{\tau}\equiv a^{\tau\mu}_ky^k_{\mu+1_i}+({\partial}_ia^{\tau\mu}_k)y^k_{\mu}\] 
in order to induce maps $d_i:M \longrightarrow M:{\bar{y} }^k_{\mu} \longrightarrow {\bar{y}}^k_{\mu+1_i}$ by residue if we use to denote the residue $Dy \longrightarrow M: y^k \longrightarrow {\bar{y}}^k$ by a bar as in algebraic geometry. However, for simplicity, we shall not write down the bar when the background will indicate clearly if we are in $Dy$ or in $M$. We have a filtration $0 \subseteq M_0 \subseteq M_1 \subseteq  ... \subseteq M_q \subseteq ... \subseteq M_{\infty}=M$ induced by that of $D$ and $d_iM_q\subseteq M_{q+1}$ (Compare to [4] and [37]).  \\

As a byproduct, the differential modules we shall consider will always be {\it finitely generated} ($k=1,...,m<\infty$) and {\it finitely presented} ($\tau=1, ... ,p<\infty$). Equivalently, introducing the {\it matrix of operators} ${\cal{D}}=(a^{\tau\mu}_kd_{\mu})$ with $m$ columns and $p$ rows, we may introduce the morphism $D^p \stackrel{{\cal{D}}}{\longrightarrow} D^m:(P_{\tau}) \longrightarrow (P_{\tau}{\Phi}^{\tau}):P \longrightarrow P\Phi=P{\cal{D}}$ over $D$ by acting with $D$ {\it on the left of these row vectors} while acting with ${\cal{D}}$ {\it on the right of these row vectors} and the {\it presentation} of $M$ is defined by the exact cokernel sequence $D^p \longrightarrow D^m \longrightarrow M \longrightarrow 0 $. It is essential to notice that the presentation only depends on $K, D$ and $\Phi$ or $ \cal{D}$, that is to say never refers to the concept of (explicit or formal) solutions. It is at this moment that we have to take into account the results of the previous section in order to understant that certain presentations will be much better than others, in particular to establish a link 
with formal integrability and involution. \\

Having in mind that $K$ is a left $D$-module with the standard action $(D,K) \longrightarrow K:(d_i,a)\longrightarrow {\partial}_ia$ and that $D$ is a bimodule over itself, {\it we have only two possible constructions}:  \\

\noindent
{\bf DEFINITION 2C.1}: We define the {\it system} $R=hom_K(M,K)=M^*$ and set $R_q=hom_K(M_q,K)=M_q^*$ as the {\it system of order} $q$. We have the {\it projective limit} $R=R_{\infty} \longrightarrow ... \longrightarrow R_q \longrightarrow ... \longrightarrow R_1 \longrightarrow R_0$. It follows that $f_q\in R_q:y^k_{\mu} \longrightarrow f^k_{\mu}\in K$ with $a^{\tau\mu}_kf^k_{\mu}=0$ defines a {\it section at order} $q$ and we may set $f_{\infty}=f\in R$ for a {\it section} of $R$. For a ground field of constants $k$, this definition has of course to do with the concept of a formal power series solution. However, for an arbitrary differential field $K$, {\it the main novelty of this new approach is that such a definition has nothing to do with the concept of a formal power series solution} ({\it care}) as illustrated in the next example.\\

\noindent
{\bf DEFINITION 2C.2}: We may define the right differential module $hom_D(M,D)$.  \\

\noindent
{\bf PROPOSITION 2C.3}: When $M$ is a left $D$-module, then $R$ is also a left $D$-module. \\

\noindent
{\it Proof}: As $D$ is generated by $K$ and $T$ as we already said, let us define:  \\
\[  (af)(m)=af(m), \hspace{4mm} \forall a\in K, \forall m\in M \]
\[ (\xi f)(m)=\xi f(m)-f(\xi m), \hspace{4mm} \forall \xi=a^id_i\in T,\forall m\in M  \]
In the operator sense, it is easy to check that $d_ia=ad_i+{\partial}_ia$ and that $\xi\eta - \eta\xi=[\xi,\eta]$ is the standard bracket of vector fields. We finally 
get $(d_if)^k_{\mu}=(d_if)(y^k_{\mu})={\partial}_if^k_{\mu}-f^k_{\mu +1_i}$ and thus recover {\it exactly} the Spencer operator of the previous section though {\it this is not evident at all}. We also get $(d_id_jf)^k_{\mu}={\partial}_{ij}f^k_{\mu}-{\partial}_if^k_{\mu+1_j}-{\partial}_jf^k_{\mu+1_i}+f^k_{\mu+1_i+1_j} \Longrightarrow d_id_j=d_jd_i, \forall i,j=1,...,n$ and thus $d_iR_{q+1}\subseteq R_q\Longrightarrow d_iR\subset R$ induces a well defined operator $R\longrightarrow T^*\otimes R:f \longrightarrow dx^i\otimes d_if$. This result has been discovered (up to sign) by Macaulay  in 1916 ([18]). For more details on the Spencer operator and its applications, the reader may look at ([22],[23]).  \\
\hspace*{12cm}   Q.E.D.  \\

We now recall the definition of the {\it extension modules} $ext_D^i(M,D)$ that we shall simply denote by $ext^i(M)$ when there cannot be any confusion. We divide the procedure into four steps that can be achieved by means of computer algebra ([29],[34-35]): \\

\noindent
$\bullet$ Construct a {\it free resolution} of $M$, say:  \\
\[   ... \longrightarrow F_i \longrightarrow ... \longrightarrow F_1 \longrightarrow F_0 \longrightarrow M \longrightarrow 0  \]

\noindent
$\bullet$ Suppress $M$ in order to obtain the {\it deleted sequence}:  \\
\[     ... \longrightarrow F_i \longrightarrow ... \longrightarrow F_1 \longrightarrow F_0 \longrightarrow 0  \hspace{11mm}  \]

\noindent
$\bullet$ Apply $hom_D(\bullet,D)$ in order to obtain the {\it dual sequence} heading backwards: \\
\[     ... \longleftarrow hom_D(F_i,D) \longleftarrow ... \longleftarrow hom_D(F_1,D) \longleftarrow hom_D(F_0,D) \longleftarrow 0    \]

\noindent
$\bullet$ Define $ext^i(M)$ to be the cohomology at $hom_D(F_i,D)$ in the dual sequence with $ext^0(M)=hom_D(M,D)$.  \\

The following nested chain of difficult propositions and theorems can be obtained, {\it even in the non-commutative case}, by combining the use of extension modules and {\it double duality} in the framework of algebraic analysis ([4],[25]). \\

\noindent
{\bf THEOREM 2C.4}: The extension modules do not depend on the resolution of $M$ used.  \\

\noindent
{\bf PROPOSITION 2C.5}: Applying $hom_D(\bullet,D)$ provides right $D$-modules that can be transformed to left $D$-modules by means of the {\it side changing functor} and vice-versa. Namely, if $N_D$ is a right $D$-module, then ${}_DN={\wedge}^nT{\otimes}_KN_D$ is the {\it converted left} $D$-module while, if ${}_DN$ is a left $D$-module, then $N_D={\wedge}^nT^*{\otimes}_K{}_DN$ is the {\it converted right} $D$-module.\\

\noindent
{\bf PROPOSITION 2C.6}: Instead of using $hom_D(\bullet,D)$ and the side changing functor in the module framework, we may use $ad$ in the operator framework. Namely, to any operator ${\cal{D}}:E \longrightarrow F$ we may associate the formal adjoint $ad({\cal{D}}):{\wedge}^nT^*\otimes F^*\longrightarrow {\wedge}^nT^*\otimes E^*$ with the useful though striking relation $rk_D(ad({\cal{D}}))=rk_D({\cal{D}})$.  \\

\noindent
{\bf PROPOSITION 2.C.7}: $ext^i(M)$ is a torsion module $\forall 1\leq i \leq n$ but $ext^0(M)=hom_D(M,D)$ may not be a torsion module.  \\

We shall say that an operator is {\it parametrizable} if it generates the CC of that operator and the next result will be essential for applications as it can be tested by means of computer algebra ([34],[35]).  \\

\noindent
{\bf THEOREM 2C.8}: An operator is {\it parametrizable} if and only if the corresponding differential module is torsion-free and double duality provides a constructive test for checking such a property ([34]). \\

\hspace{1cm}

\noindent
{\bf 3) APPLICATIONS}  \\

\noindent
{\bf A) MINKOWSKI METRIC}: 

If $n=4$ and $\omega \in S_2T^*$ is the non-degenerate Minkowski metric, the corresponding Lie operator is $\xi \in T \rightarrow \Omega \equiv {\cal{D}}\xi={\cal{L}}(\xi)\omega \in S_2T^*$ where ${\cal{L}}$ is the standard Lie derivative for tensors and we have to study the corresponding system $R_1 \subset J_1(T)$ of infinitesimal Lie equations. However, this system is finite type with $g_2=0$ but $g_1\subset T^*\otimes T$ is {\it not} $2$-acyclic and the CC are homogeneous of order $2$, a result leading to the well known finite length differential sequence where the order of an operator has been indicated under its arrow: \\
\[  0 \rightarrow \Theta  \rightarrow 4 \underset{1}{\stackrel{Killing}{\longrightarrow}} 10 \underset{2}{\stackrel{Riemann}{\longrightarrow}} 20 \underset{1}{\stackrel{Bianchi}{\longrightarrow}} 20 \underset{1}{\longrightarrow} 6  \rightarrow 0  \]
In arbitrary dimension, we have successively:  \\
\[  n \rightarrow  n(n+1)/2 \rightarrow n^2(n^2-1)/12 \rightarrow n^2(n^2-1)(n-2)/24 \rightarrow ...    \]
or, introducing the Spencer $\delta$-cohomology:  \\
\[   T \rightarrow S_2T^* \rightarrow H^2(g_1) \rightarrow H^3(g_1) \rightarrow ...  \]
However, this sequence is {\it not} canonical and we have to use the involutive system $R_2\subset J_2(T)$ in the following {\it Fundamental Diagram I} relating the (upper) {\it canonical Spencer sequence} to the (lower) {\it canonical Janet sequence}, a result first found exactly 40 years ago in ([20]) that only depends on the left commutative square ${\cal{D}}={\Phi}_0\circ j_q$ when one has an involutive system $R_q\subseteq J_q(E)$ over $E$ with $dim(X)=n$:  \\

\[   SPENCER   \]
 \[ \footnotesize  \begin{array}{rcccccccccccccl}
 &&&&& 0 &&0&&0&&  &  &0&  \\
 &&&&& \downarrow && \downarrow && \downarrow & &  &   & \downarrow &  \\
  & 0& \rightarrow& \Theta &\stackrel{j_q}{\rightarrow}&C_0 &\stackrel{D_1}{\rightarrow}& C_1 &\stackrel{D_2}{\rightarrow} & C_2 &\stackrel{D_3}{\rightarrow}& ... &\stackrel{D_n}{\rightarrow}& C_n &\rightarrow 0 \\
  &&&&& \downarrow & & \downarrow & & \downarrow & &  &&\downarrow &     \\
   & 0 & \rightarrow & E & \stackrel{j_q}{\rightarrow} & C_0(E) & \stackrel{D_1}{\rightarrow} & C_1(E) &\stackrel{D_2}{\rightarrow} & C_2(E) &\stackrel{D_3}{\rightarrow} & ... &\stackrel{D_n}{\rightarrow} & C_n(E) &   \rightarrow 0 \\
   & & & \parallel && \hspace{6mm}\downarrow  {\Phi}_0& & \hspace{6mm}\downarrow  {\Phi}_1& & \hspace{6mm}\downarrow {\Phi}_2 & &  & & \hspace{6mm} \downarrow  {\Phi}_n & \\
   0 \rightarrow & \Theta &\rightarrow & E & \stackrel{\cal{D}}{\rightarrow} & F_0 & \stackrel{{\cal{D}}_1}{\rightarrow} & F_1 & \stackrel{{\cal{D}}_2}{\rightarrow} & F_2 & \stackrel{{\cal{D}}_3}{\rightarrow} & ... &\stackrel{{\cal{D}}_n}{\rightarrow} & F_n & \rightarrow  0 \\
   &&&&& \downarrow & & \downarrow & & \downarrow & &  &  &\downarrow &   \\
   &&&&& 0 && 0 && 0 && &&0 &  
   \end{array}     \]
\[   JANET   \]
Setting $n=4,q=2,E=T,C_r={\wedge}^rT^*\otimes R_2$, the first order {\it first Spencer operator} $D_1$ is defined by the system $R_3\subset J_1(R_2)$ and the epimorphisms ${\Phi}_1 , ..., {\Phi}_4$ are induced by the epimorphism ${\Phi}_0$. The corresponding fiber dimensions are indicated in the next diagram where $dim (R_2)=dim(T) + dim(g_1)=4+6=10$: \\
 \[ \small  \begin{array}{rcccccccccccccl}
 &&&&& 0 &&0&&0&& 0 &  &0&  \\
 &&&&& \downarrow && \downarrow && \downarrow & & \downarrow &   & \downarrow &  \\
  & 0& \rightarrow& \Theta &\stackrel{j_2}{\rightarrow}&10 &\stackrel{D_1}{\rightarrow}& 40 &\stackrel{D_2}{\rightarrow} & 60 &\stackrel{D_3}{\rightarrow}&40 &\stackrel{D_4}{\rightarrow}& 10 &\rightarrow 0 \\
  &&&&& \downarrow & & \downarrow & & \downarrow & & \downarrow &&\downarrow &     \\
   & 0 & \rightarrow & 4 & \stackrel{j_2}{\rightarrow} & 60 & \stackrel{D_1}{\rightarrow} & 160 &\stackrel{D_2}{\rightarrow} & 180 &\stackrel{D_3}{\rightarrow} & 96 &\stackrel{D_4}{\rightarrow} &20&   \rightarrow 0 \\
   & & & \parallel && \downarrow  & & \downarrow  & & \downarrow  & & \downarrow & & \downarrow  & \\
   0 \rightarrow & \Theta &\rightarrow & 4 & \stackrel{\cal{D}}{\rightarrow} & 50 & \stackrel{{\cal{D}}_1}{\rightarrow} & 120 & \stackrel{{\cal{D}}_2}{\rightarrow} & 120 & \stackrel{{\cal{D}}_3}{\rightarrow} & 56 &\stackrel{{\cal{D}}_4}{\rightarrow} & 10 & \rightarrow  0 \\
   &&&&& \downarrow & & \downarrow & & \downarrow & & \downarrow &  &\downarrow &   \\
   &&&&& 0 && 0 && 0 &&0 &&0 &  
   \end{array}     \]
We recall that, apart from $j_2$ and ${\cal{D}}={\Phi}_0\circ j_2$ which have order $2$, then all the other operators have order $1$.
Of course, the reader may imagine easily that the author of this paper avoided computer algebra by using the following specific procedure:  \\
\noindent
1)First write down the Spencer sequence at once, keeping in mind that it is locally isomorphic to the tensor product of the Poincar\' e sequence:  \\
\[   {\wedge}^0T^* \stackrel{d}{\rightarrow} {\wedge}^1T^* \stackrel{d}{\rightarrow}  {\wedge}^2T^* \stackrel{d}{\rightarrow} {\wedge}^3T^* \stackrel{d}{\rightarrow} {\wedge}^4T^* \rightarrow  0  \]     
\[1 \stackrel{d}{\rightarrow}  4   \stackrel{d}{\rightarrow}6  \stackrel{d}{\rightarrow}  4  \stackrel{d}{\rightarrow} 1 \rightarrow 0\]
by a Lie algebra of dimension $4+6=10$ and is thus locally exact .  \\
\noindent
2) Second, write down the central sequence just using some combinatorics on the Janet tabular for the trivially involutve first operator $j_2$ (see [20], p 157). \\ 
\noindent
3) Then, obtain the Janet sequence by quotient.  \\
\noindent
4) Finally, use the vanishing of the three Euler-Poincar\'e characteristics for checking the exactness of all these numbers, namely:  \\
\[   10 - 40 +60 - 40 + 4=0, \hspace{3mm}  4 - 60 +160 - 180 + 96 - 20 =0, \hspace{3mm} 4 - 50 + 120 - 120 + 56 - 10 = 0   \]

\vspace{3cm}

\noindent
{\bf B) SCHWARZSCHILD METRIC}: 

In the Boyer-Lindquist coordinates $(t,r,\theta, \phi)=(x^0,x^1,x^2,x^3)$, we consider the Schwarzschild metric $\omega= A(r)dt^2-(1/A(r))dr^2 - r^2d{\theta}^2 -r^2{sin}^2(\theta)d{\phi}^2$ and $\xi={\xi}^id_i \in T$, let us  introduce ${\xi}_i={\omega}_{ri}{\xi}^r$ with the $4$ {\it formal derivatives} $(d_0=d_t, d_1=d_r,d_2= d_{\theta}, d_3=d_{\phi})$. With speed of light $c=1$ and $A=1-\frac{m}{r}$ where $m$ is a constant, the metric can be written in the diagonal form:  \\
\[ \left(    \begin{array}{cccc}
A & 0  &  0  &  0   \\
0 & -1/A & 0 & 0 \\
0 & 0 & -r^2 & 0 \\
0 & 0 & 0 & -r^2sin^2(\theta)
\end{array}  \right)   \]
Using the notations of differential modules theory, consider 
the Killing equations:   \\
\[ \Omega \equiv {\cal{L}}(\xi)\omega=0 \hspace{1cm} \Leftrightarrow \hspace{1cm} {\Omega}_{ij}\equiv d_i{\xi}_j +d_j{\xi}_i  - 2 {\gamma}^r_{ij}{\xi}_r=0\]
where we have introduced the Christoffel symbols $\gamma$ through he standard Levi-Civita isomorphism $j_1(\omega)\simeq (\omega, \gamma)$ while  setting $A'={\partial}_rA$ in the differential field $K$ of coefficients ([8], p 87). 
As in the later Macaulay and Janet examples and in order to avoid any further confusion between sections and derivatives, we shall use the sectional point of view and rewrite the previous equations in the symbolic form $\Omega \equiv  L({\xi}_1)\omega \in S_2T^* $ where $L$ is the {\it formal Lie derivative}:  \\ 
\[  \left\{ \begin{array}{rcl}
\frac{1}{2}{\Omega}_{33} &  \equiv  &  {\xi}_{3,3} + sin(\theta)cos(\theta){\xi}_2 +rAsin^2(\theta) {\xi}_1 =0  \\
{\Omega}_{32  } &  \equiv  &  {\xi}_ {2,3}+ {\xi}_ {3,2}  - 2 cot(\theta)   {\xi}_ 3   =0   \\
{\Omega}_{31  } &  \equiv  &  {\xi}_ {1,3}+ {\xi}_{3,1}   -  \frac{2}{r}  {\xi}_ 3   =0  \\
{\Omega}_{30  } &  \equiv  &  {\xi}_ {0,3}+ {\xi}_ {3,0}    =0   \\
\frac{1}{2}{\Omega}_{22} &  \equiv  &  {\xi}_{2,2} + r A {\xi}_1 =0  \\
{\Omega}_{21  } &  \equiv  &  {\xi}_{1,2}+ {\xi}_{2,1}  - \frac{2}{r}   {\xi}_ 2   =0   \\
{\Omega}_{20  } &  \equiv  &  {\xi}_{0,2}+ {\xi}_{2,0}     =0  \\
\frac{1}{2}{\Omega}_{11}  &  \equiv  &  {\xi}_{1,1} + \frac{A'}{2A}{\xi}_1=0   \\
{\Omega}_{10}  &  \equiv  &  {\xi}_{0,1} + {\xi}_{1,0} - \frac{A'}{A}{\xi}_0 =0  \\
\frac{1}{2}{\Omega}_{00} &  \equiv  &    {\xi}_{0,0}  - \frac{AA'}{2}{\xi}_1=0
\end{array} \right.  \]

Though this system $R_1\subset J_1(T)$ has 4 equations of class $3$, 3 equations of class $2$, 2 equations of class $1$ and 1 equation of class $0$, it is far from being involutive because it is finite type with second symbol $g_2=0$ defined by the 40 equations $v^k_{ij}=0$ in the initial coordinates. From the symetry, it is clear that such a system has at least 4 solutions, namely the time translation ${\partial}_t \leftrightarrow {\xi}^0=1 \Leftrightarrow {\xi}_0= A$ and, using cartesian coordinates $(t,x,y,z)$, the 3 space rotations $y{\partial}_z-z{\partial}_y, z{\partial}_x - x{\partial}_z, x{\partial}_y - y{\partial}_x$. \\

We may also write the Schwarzschild metric in cartesian coordinates as:ÊÊ\\
\[   \omega=A(r)dt^2 + (1-\frac{1}{A(r)})dr^2 - (dx^2+dy^2+dz^2), \hspace{1cm}  rdr=xdx+ydy+zdz  \]
and notice that the $3\times 3$ matrix of components of the three rotations has rank equal to $2$, a result leading surely, before doing any computation to the existence of {\it one and only one} zero order Killing equation $x{\xi}^x+y{\xi}^y+z{\xi}^z=0$.  Such a result also amounts to say that the spatial projection of any Killing vector on the radial spatial unit vector $(x^1/r,x^2/r,x^3/r)$ vanishes beause $r$ must stay invariant. \\

All the Christoffel symbols vanish but:  \\
\[  {\gamma}^0_{01}=\frac{m}{2r^2A}, \,\, {\gamma}^1_{00}=\frac{mA}{2r^2}, \,\, {\gamma}^1_{11}= - \frac{m}{2r^2A}, \,\,
 {\gamma}^1_{22}= - rA, \,\, {\gamma}^1_{33}= - rAsin^2(\theta), \] 
\[ {\gamma}^2_{12}=\frac{1}{r}, \,\, {\gamma}^2_{33}= - cos(\theta)sin(\theta), \,\, {\gamma}^3_{13}=\frac{1}{r}, \,\, 
{\gamma}^3_{23}=cot(\theta)  \]
As for their linearization, setting ${\gamma}_{r,ij}={\omega}_{kr}{\gamma}^k_{ij}= \frac{1}{2}({\partial}_i{\omega}_{rj} + {\partial}_j{\omega}_{ir}-{\partial}_r{\omega}_{ij})$, we get:  \\
\[  {\Gamma}_{r,ij}= {\omega}_{kr}{\Gamma}^k_{ij} + {\gamma}^k_{ij}{\Omega}_{kr}=\frac{1}{2}(d_i{\Omega}_{rj}+d_j{\Omega}_{ir} - d_r{\Omega}_{ij})  \]
Now, caring only about the last three equations, we obtain formally {\it modulo} $\Omega$:  \\
\[  \begin{array}{rcl}
d_1{\Omega}_{01}={\xi}_{0,11} +{\xi}_{1,01} - d_1(\frac{A'}{A}{\xi}_0) & = & {\xi}_{0,11} - \frac{A'}{2A}{\xi}_{1,0} -(\frac{A"}{A}- \frac{(A')^2}{A^2}){\xi}_0
  - \frac{A'}{A}{\xi}_{0,1}  \\
     &  =  &  {\xi}_{0,11} + \frac{A'}{2A}{\xi}_{0,1}- \frac{(A')^2}{2A^2}{\xi}_0 - (\frac{A"}{A}- \frac{(A')^2}{A^2}){\xi}_0
  - \frac{A'}{A}{\xi}_{0,1}  \\
   & = & {\xi}_{0,11}  + \frac{A'}{2A} \,{\xi}_{1,0} - \frac{A"}{A} {\xi}_0 
     \end{array}   \]                             
and similarly in $R_2 $ with lower indices as usual:   \\
\[  \begin{array}{l}
  {\xi}_{0,11}  + \frac{A'}{2A} \,{\xi}_{1,0} - \frac{A"}{A} {\xi}_0 =0  \\
  {\xi}_{0,01}- (\frac{AA"}{2} + \frac{(A')^2}{4}){\xi}_1=0  \\
  {\xi}_{0,00}  - \frac{AA'}{2}\, {\xi}_{1,0} =0  \\ 
  {\xi}_{1,00} + (\frac{AA"}{2} - \frac{{A'}^2}{4}) \, {\xi}_1=0  \\ 
  {\xi}_{1,01} + \frac{A'}{2A} {\xi}_{1,0} =0  \\
  {\xi}_{1,11}+ (\frac{A"}{2A} - \frac{3(A')^2}{4A^2}){\xi}_1=0  
\end{array}  \]
It follows that we obtain in $R_3$ and thus finally in $R^{(2)}_1$ for calculating \fbox{${\rho}_{01,01}$}:  \\
\[  \begin{array}{l}
+{\xi}_{1,001} + \frac{A'}{2A} \, {\xi}_{1,00} =0  \\
- {\xi}_{1,001} - (\frac{AA"}{2}-\frac{(A')^2}{4})' \, {\xi}_1 - (\frac{AA"}{2} -\frac{(A')^2}{4}) \, {\xi}_{1,1}=0  \\
-\frac{A'}{2A} \, {\xi}_{1,00} - \frac{A'}{2A}(\frac{AA"}{2} - \frac{(A')^2}{4}) {\xi}_1 =0   \\
(\frac{AA"}{2} - \frac{(A')^2}{4}){\xi}_{1,1} + \frac{A'}{2A}(\frac{AA"}{2} - \frac{(A')^2}{4}){\xi}_1=0
\end{array}  \]
Summing these four prolongations, we get ${\xi}^1d_1(\frac{AA"}{2}-\frac{(A')^2}{4})=0 \Rightarrow $ \fbox{${\xi}_1=0$} because $A=1 - \frac{m}{r} \Rightarrow A+rA'=1\Rightarrow 2A' +rA"=0$. 
Similarly, we could have obtained:   \\
\[  d_1{\xi}_{1,00} - d_0 {\xi}_{1,01} = - (\frac{AA"}{2} - \frac{(A')^2}{4})' {\xi}_1=0   \]

Using the relation $A+rA'=1$, we have also successively for calculating \fbox{${\rho}_{02,01}$}:  \\
\[d_1{\Omega}_{02} + d_0{\Omega}_{12} - d_2{\Omega}_{01}=2 {\xi}_{2,01} - (\frac{2}{r}+ \frac{A'}{A}) {\xi}_{2,0} \Rightarrow {\xi}_{2,01}\\
  = (\frac{1}{r}+ \frac{A'}{2A}) {\xi}_{2,0}   \]
\[  {\xi}_{2,00}= - {\xi}_{0,20}=-\frac{AA'}{2}{\xi}_{1,2}, \hspace{5mm}  {\xi}_{0,12}= - (\frac{1}{r}+ \frac{A'}{2A}) {\xi}_{2,0} \]
\[   \begin{array}{rcl}
d_1({\xi}_{2,00}) - d_0({\xi}_{2,01})  & = & - \frac{(AA')'}{2}{\xi}_{1,2} - \frac{AA'}{2} {\xi}_{1,12}-(\frac{1}{r}+ \frac{A'}{2A}){\xi}_{2,00}     \\
                   &  =  & - (\frac{AA"}{2} + \frac{(A')^2}{4}){\xi}_{1,2}+ \frac{AA'}{2}(\frac{1}{r} + \frac{A'}{2A}){\xi}_{1,2}  \\
                       &  =  &  (\frac{AA'}{2r}- \frac{AA"}{2}){\xi}_{1,2}   \\
                         &  =  &  \frac{3AA'}{2r}{\xi}_{1,2}  
                        
\end{array}   \]
that is to say \fbox{${\xi}_{1,2}=0$}. However, we have for calculating \fbox{${\rho}_{11,01}$}:  \\
\[  \begin{array}{rcl}
 d_1({\xi}_{1,01}) - d_0({\xi}_{1,11})  & =  &   d_1(-\frac{A'}{2A}{\xi}_{1,0}) - ( \frac{3(A')^2}{4A^2} -\frac{A"}{2A}) {\xi}_{1,0 }    \\                
   & = &  -\frac{A'}{2A}{\xi}_{1,01} - (\frac{A"}{2A}- \frac{(A')^2}{2A^2}){\xi}_{1,0}- ( \frac{3(A')^2}{4A^2} -\frac{A"}{2A}) {\xi}_{1,0 }\\
  &  =  & (\frac{(A')^2}{4A^2}  + \frac{(A')^2}{2A^2}  - \frac{3(A')^2}{4A^2} ){\xi}_{1,0}\\
  &  =  & 0
\end{array}  \]
and such an approach does not bring ${\xi}_{1,0}=0$ for sections, even if it {\it must be surely true} for solutions. We may also notice that:
\[ d_1{\Omega}_{02} + d_2{\Omega}_{01} - d_0{\Omega}_{12}=2 {\xi_{0,12}- \frac{A'}{A}{\xi}_{0,2}}+ \frac{2}{r}{\xi}_{2,0}\]
\[  \Rightarrow  {\xi}_{0,12}=(\frac{A'}{2A}+ \frac{1}{r}){\xi}_{0,2}   \]
\[ d_0{\Omega}_{12} + d_2{\Omega}_{01}-d_1{\Omega}_{02}= 2 {\xi}_{1,02} -\frac{2}{r}{\xi}_{2,0} - \frac{A'}{A}{\xi}_{0,2}  \]
\[  \Rightarrow  {\xi}_{1,02}=(\frac{A'}{2A}- \frac{1}{r}){\xi}_{0,2}  \]
\[      {\xi}_{0,11}= \frac{A"}{A}{\xi}_0 - \frac{A'}{2A}{\xi}_{1,0}  \]
Studying the component \fbox{${\rho}_{01,12}$}, we obtain successively:  \\
 \[   \begin{array}{rcl}
d_2{\xi}_{0,11}- d_1{\xi}_{0,12} &  =  & \frac{A"}{A}{\xi}_{0,2} - \frac{A'}{2A} {\xi}_{1,02}- (\frac{A'}{2A}+\frac{1}{r})' {\xi}_{0,2}-
(\frac{A'}{2A} + \frac{1}{r}){\xi}_{0,12}  \\
 & = & [ \frac{A"}{A}- \frac{A'}{2A}(\frac{A'}{2A}- \frac{1}{r})- (\frac{A'}{2A}+ \frac{1}{r})' - (\frac{A'}{2A}+ \frac{1}{r})^2] {\xi}_{0,2} \\
 &  =  &  - \frac{2A'}{rA} {\xi}_{0,2}
\end{array}  \]
and thus \fbox{${\xi}_{0,2}=0 $}.  We also have:   \\
\[  d_3{\Omega}_{01} + d_1{\Omega}_{03} - d_0{\Omega}_{13}= 2 {\xi}_{0,13}- \frac{A'}{A}{\xi}_{0,3} +\frac{2}{r} {\xi}_{3,0}  \]
\[  \Rightarrow   {\xi}_{0,13}= (\frac{A'}{2A} + \frac{1}{r}){\xi}_{0,3}     \]
\[ d_0{\Omega}_{13} + d_3{\Omega}_{01} - d_1{\Omega}_{03}= 2 {\xi}_{1,03} - \frac{2}{r}{\xi}_{3,0} -\frac{A'}{A} {\xi}_{0,3}  \]
\[ \Rightarrow   {\xi}_{1,03} = (\frac{A'}{2A} - \frac{1}{r}) {\xi}_{0,3}  \]
Studying the component \fbox{${\rho}_{01,13}$}, we obtain successively:  \\
\[ {\xi}_{0,11}= \frac{A"}{A}{\xi}_0 - \frac{A'}{2A} {\xi}_{1,0}      \]
\[  d_3{\Omega}_{01} + d_1{\Omega}_{03} - d_0{\Omega}_{13}=2 {\xi}_{0,13}- \frac{A'}{A} {\xi}_{0,3}  \]
\[  d_0{\Omega}_{13} + d_3 {\Omega}_{01} - d_1 {\Omega}_{0,3}= 2 {\xi}_{1,03}- \frac{2}{r}{\xi}_{3,0}- \frac{A'}{A}{\xi}_{0,3}  \]
\[  \Rightarrow   {\xi}_{1,03}= (\frac{A'}{2A} - \frac{1}{r}) {\xi}_{0,3}  \]
\[   \begin{array}{rcl}
d_3{\xi}_{0,11} - d_1 {\xi}_{0,13} & = & \frac{A"}{A}{\xi}_{0,3}- \frac{A'}{2A}{\xi}_{1,03} - (\frac{A'}{2A} +\frac{1}{r})'{\xi}_{0,3}- (\frac{A'}{2A}+ \frac{1}{r}){\xi}_{0,13}  \\
&= & [ \frac{A"}{A} - \frac{A'}{2A}( \frac{A'}{2A}- \frac{1}{r}) - (\frac{A'}{2A} +\frac{1}{r})' - (\frac{A'}{2A}+ \frac{1}{r})^2]{\xi}_{0,3} \\
    & = &   - \frac{3A'}{2rA}{\xi}_{0,3}
\end{array}  \]
and thus \fbox{${\xi}_{0,3}=0$}. We also have:  \\
\[  d_1{\Omega}_{03}+ d_0{\Omega}_{13} - d_3{\Omega}_{01}= 2 {\xi}_{3,01} - \frac{2}{r}{\xi}_{3,0} - \frac{A'}{A}{\xi}_{0,3}  \]
\[ \Rightarrow {\xi}_{3,01}=  (\frac{A'}{2A} - \frac{1}{r}){\xi}_{0,3}   , \hspace{1cm} {\xi}_{3,00}= - {\xi}_{0,03}= - \frac{AA'}{2}{\xi}_{1,3} \]
Studying \fbox{${\rho}_{03,01}$}, we obtain successively:  \\
\[  \begin{array}{rcl}
 d_0{\xi}_{3,01}  -d_1{\xi}_{3,00}  & = & (\frac{A'}{2A}-\frac{1}{r}){\xi}_{0,03} + (\frac{AA'}{2})' {\xi}_{1,3} + \frac{AA'}{2} {\xi}_{1,13}  \\
 &  =  &  [ \frac{AA'}{2}(\frac{A'}{2A}- \frac{1}{r}) + (\frac{AA'}{2})' - \frac{(A')^2}{4} ] {\xi}_{1,3}   \\
  &  =  &  ((\frac{AA'}{2})' - \frac{AA'}{r}){\xi}_{1,3} 
\end{array}   \]
and thus \fbox{${\xi}_{1,3}=0$}. We also have:  \\
\[  d_0{\Omega}_{12}+d_1{\Omega}_{02}-d_2{\Omega}_{01}= 2 {\xi}_{2,01} - \frac{2}{r}{\xi}_{2,0}- (\frac{A'}{A}){\xi}_{2,0}=0  \]
\[   \Rightarrow    {\xi}_{2,01} = ( \frac{1}{r}+ \frac{A'}{2A}){\xi}_{2,0} \]
Finally, studying the component \fbox{${\rho}_{21,02}$} when $(rA)'=1$, we have successively:  \\
\[  \begin{array}{rcl}
d_0({\xi}_{2,12}) -  d_2({\xi}_{2,01}) & = & d_0(-{\xi}_1-rA{\xi}_{1,1})  -(\frac{1}{r}+ \frac{A'}{2A}){\xi}_{2,02}  \\
 & = & \frac{rA'}{2} {\xi}_{1,0} + rA(\frac{1}{r} + \frac{A'}{2A}){\xi}_{1,0}-{\xi}_{1,0}  \\
& = &   (A+rA' - 1){\xi}_{1,0}  \\
  &  =  &   0
   \end{array}   \]
and thus ${\xi}_{1,0}=0$ {\it cannot be obtained}. As we already proved that we had ${\xi}_1=0$ and thus  ${\xi}_{1,1}=0$ but also ${\xi}_{1,2}=0, {\xi}_{0,2}=0 ,{\xi}_{1,3}=0, {\xi}_{0,3}=0$, we have therefore obtained $10+5=15$ linearly independent first order equations after {\it only} $2$ prolongations that can also be obtained by computer algebra in a rather "{\it brute}" way. \\
Accordingly, {\it one needs one more prolongation in order to obtain} \fbox{${\xi}_{1,0}=0$} {\it from} ${\xi}_1=0$ {\it by setting} ${\xi}_{1,0}\equiv d_0{\xi}_1$ {\it formally}, {\it in such a way that} $d_0{\xi}_1-{\xi}_{1,0}=0$ {\it is the identity} $0=0$, as in Example 2A.9. \\

\noindent
{\bf REMARK 3B.1}: We present an alternative approach for finding the same results and illustrate it on three cases. First of all we obtain easily:  \\
\[  {\xi}^0_0 + \frac{A'}{2A}{\xi}^1=0, {\xi}^1_1- \frac{A'}{2A}{\xi}^1=0, {\xi}^2_2 +\frac{1}{r}{\xi}^1=0, {\xi}^3_3 + cot(\theta){\xi}^2 + \frac{1}{r} {\xi}^1=0 \]
 with ${\xi}_0=A{\xi}^0, {\xi}_1=-\frac{1}{A}{\xi}^1, {\xi}_2=-r^2{\xi}^2, {\xi}_3= - r^2sin^2({\theta}){\xi}^3$. Then, using $r$ as a summation index, we shall see that we have {\it in general}:  \\
\[  R_{kl,ij}\equiv {\rho}_{rl,ij}{\xi}^r_k +{\rho}_{kr,ij}{\xi}^r_l+{\rho}_{kl,rj}{\xi}^r_i + {\rho}_{kl,ir}{\xi}^r_j + {\xi}^r {\partial}_r 
{\rho}_{kl,ij}\neq 0  \]
but will notice that ${\rho}_{ij}=0 \Rightarrow R_{ij}\equiv {\rho}_{rj}{\xi}^r_i + {\rho}_{ir}{\xi}^r_j +{\xi}^r{\partial}_r{\rho}_{ij}=0  $.\\
The $6$ non-zero components of the Weyl tensor are:  \\
\[   \begin{array}{lll}
  {\rho}_{01,01}= + \frac{m}{r^3},& \,\,{\rho}_{02,02}= - \frac{m\,A}{2r}, &\,\,{\rho}_{03,03}= - \frac{m\,A\,sin^2(\theta)}{2r}  \\
  {\rho}_{12,12}= + \frac{m}{2r\,A},& \,\,{\rho}_{13,13}= + \frac{m\,sin^2(\theta)}{2r\,A},& \,\, {\rho}_{23,23}= - m\, r \, sin^2(\theta)  
\end{array}   \]
We obtain in particular:  \\
\[  R_{01,01}\equiv 2{\rho}_{01,01} ({\xi}^0_0 + {\xi}^1_1) + {\xi}^r{\partial}_r({\rho}_{01,01})={\xi}^1{\partial}_1{\rho}_{01,01}=0   \Rightarrow {\xi}^1=0 \]
and similarly:  \\
\[  R_{01,02} \equiv {\rho}_{01,01}{\xi}^1_2 + {\rho}_{02,02}{\xi}^2_1 +{\xi}^r{\partial}_r{\rho}_{01,02}= \frac{mA}{2r^3}( 3{\xi}_{1,2} - {\Omega}_{12})=0 \Rightarrow {\xi}^1_2=0 \]
\[  R_{02,12}\equiv {\rho}_{12,12}{\xi}^1_0+{\rho}_{02,02}{\xi}^0_1+{\xi}^r{\partial}_r{\rho}_{02,12}= - \frac{m}{2r}{\Omega}_{01}=0 \]
\[ \begin{array}{rcl}
R_{23,12} & = & {\rho}_{21,12}{\xi}^1_3 + {\rho}_{23,32}{\xi}^3_1  + {\xi}^r{\partial}_r{\rho}_{23,12}  \\
                & = &  -  \frac{m}{2rA}{\xi}^1_3 +  mrsin^2(\theta){\xi}^3_1  \\
                & = &  \frac{m}{2r}{\xi}_{1,3} -  \frac{m}{r}({\xi}_{3,1}- \frac{2}{r}{\xi}_3)  \\
                & = &  \frac{m}{2r}{\xi}_{1,3} - \frac{m}{r} ( {\Omega}_{13} - {\xi}_{1,3})  \\
                & = &  \frac{3m}{2r}{\xi}_{1,3} - \frac{m}{r} {\Omega}_{13}  
\end{array}  \]
\[  R_{01,03}= - \frac{3mA}{2r^3}{\xi}_{1,3} +\frac{mA}{2r^3}{\Omega}_{13} = - \frac{A}{r^2}R_{23,12} - \frac{mA}{2r^3}{\Omega}_{13}    \]
and so on, as a way to avoid using computer algebra. However, the main consequence of this remark is to explain the existence of the $15$ second order CC. Indeed, denoting by "$\sim$" a linear proportional dependence {\it modulo} $\Omega$, we have the three cases:  \\
\[  R_{01,01} \sim R_{02,02}\sim R_{03,03} \sim R_{12,12} \sim R_{13,13} \sim R_{23,23} \rightarrow 
{\xi}_1=0 \]
\[ R_{01,02} \sim R_{13,23} \rightarrow {\xi}_{1,2}=0\]
\[  R_{01,03} \sim R_{12,23} \rightarrow {\xi}_{1,3}=0  \]
\[  R_{01,12} \sim R_{03,23} \rightarrow {\xi}_{0,2}=0  \]
\[ R_{01,13} \sim  R_{02,23}  \rightarrow {\xi}_{0,3}=0  \]
\[  R_{01,23} \rightarrow 0, R_{02,03} \rightarrow 0, R_{02,12} \rightarrow 0, R_{02,13} \rightarrow 0, R_{03,13} \rightarrow  0,  R_{12,13} \rightarrow 0  \]
as a way to obtain the $5$ equalities to zero on the right and thus a total of $20-5=15$ second order CC obtained by elimination. However, the present partition $15= 5 + 4 + 6$ is quite different from the partition $15=10 + 5$ used by the authors quoted in the Introduction which is obtained by taking into account the vanishing assumption of the $10$ components of the Ricci tensor. As such a result questions once more the mathematical foundations of general relativity, in particular the existence of gravitational waves, we provide a few additional technical comments.\\

The main point is a tricky formula which is not evident at all. Indeed, using the well known properties of the Lie derivative, we have the following tensors and their linearizations:  \\
\[  {\rho}_{kl,ij}={\omega}_{kr}{\rho}^r_{l,ij} \,\,\,  \Rightarrow   \,\,\, R_{kl,ij}=(L({\xi}_1)\rho)_{kl,ij}= {\omega}_{kr}R^r_{l,ij} + {\rho}^r_{l,ij} {\Omega}_{kr}\,\, \Rightarrow \,\, {\omega}^{rs}R_{ri,sj} = R_{ij} + {\omega}^{rs}{\rho}^t_{i,rj}{\Omega}_{st}\] 
\[  {\rho}_{ij}={\rho}^r_{i,rj}\,\,\,  \Rightarrow  \,\,\,  R_{ij} =(L({\xi}_1)\rho)_{ij}= R^r_{i,rj}\neq {\omega}^{rs} R_{ri,sj}  \]
We prove these results using local coordinates and the formal Lie derivative obtained while replacing $j_1(\xi)$ by ${\xi}_1$ (See [20],[22],[23] for details). First of all, from the tensorial property of the Riemann tensor and the Killing equations ${\Omega}_{us}= {\omega}_{ku}{\xi}^k_s + {\omega}_{ks}{\xi}^k_u + {\xi}^r {\partial}_r{\omega}_{us}$, we have:  \\
\[  R^k_{l,ij}= (L({\xi}_1)\rho)^k_{l,ij}= - {\rho}^s_{l,ij}{\xi}^k_s + {\rho}^k_{s,ij}{\xi}^s_l + {\rho}^k_{l,sj}{\xi}^s_i + {\rho}^k_{l,is}{\xi}^s_j 
+ {\xi}^r{\partial}_r{\rho}^k_{l,ij}  \]
\[  \begin{array}{rcl}
{\omega}_{ku}( - {\rho}^s_{v,ij} {\xi}^k_s + {\xi}^r{\partial}_r{\rho}^s_{v,ij}) & =  & {\rho}^s_{v,ij}{\omega}_{ks}{\xi}^k_u + ({\xi}^r{\partial}_r{\omega}_{us}){\rho^s_{v,ij}  + {\omega}_{ku}{\xi}^r  {\partial}_r{\rho}^k_{v,ij}} -{\rho}^s_{v,ij}{\Omega}_{su} \\
  &  =  &  {\rho}_{sv,ij}{\xi}^s_u + {\xi}^r{\partial}_r  {\rho}_{uv,ij}  -{\rho}^s_{v,ij}{\Omega}_{su}                       
\end{array}  \]
and thus ${\omega}_{ku}R^k_{v,ij}= R_{uv,ij} - {\rho}^s_{v,ij}{\Omega}_{su}$.
\[  {\rho}^1_{1,11}= 0 \,\, \Rightarrow {\rho}_{11}= {\rho}^0_{1,01}+ {\rho}^2_{1,21} + {\rho}^3_{1,31}= \frac{1}{A}{\rho}_{01,01} - \frac{1}{r^2}{\rho}_{12,12} - \frac{1}{r^2sin^2(\theta)} {\rho}_{13,13} =0  \]
We have for example, {\it in this particular case}:   \\
\[ {\rho}^0_{1,01}= \frac{1}{A} {\rho}_{01,01}=\frac{m}{r^3A} \Rightarrow R^0_{1,01} = ({\cal{L}}(\xi)\rho)^0_{1,01}=2 {\rho}^0_{1,01}{\xi}^1_1+ {\xi}^1 {\partial}_1{\rho}^0_{1,01} = \frac{2m}{r^3A}{\xi}^1_1+ {\xi}^1{\partial}_1( \frac{m}{r^3A})  \]
where use only $R_{01,01}$ allowing to get ${\xi}_1=0$ in the previous list, but we have also: \\
\[  {\omega}^{rs}R_{r1,s2}= \frac{1}{A}R_{01,02} - \frac{1}{r^2sin^2(\theta)} R_{31,32}=- \frac{m}{2r^3}{\Omega}_{12}=R_{12}+{\omega}^{11}{\rho}^2_{1,12}{\Omega}_{12} \, \Rightarrow  \, R_{12}=0       \]
where we have to use $R_{01,02}$ and $R_{13,23}$ allowing to get ${\xi}_{1,2}=0$ in the previous list with:   \\
\[    R^0_{1,02}  =   - \frac{3m}{2r^3}{\xi}_{1,2} + \frac{m}{2r^3}{\Omega}_{12}, \,\,  \,\, 
R^3_{1,32}  =   + \frac{3m}{2r^3} {\xi}_{1,2} -  \frac{m}{r^3}{\Omega}_{12}  \]    
and thus:   \\
\[ {\omega}^{rs}{\rho}^t_{1,r2}{\Omega}_{st}= -{\omega}^{11}{\omega}^{22}{\rho}_{12,12}{\Omega}_{12}= - \frac{m}{2r^3}{\Omega}_{12}  \,\,\,  \Rightarrow  \,\,\, R_{12}= R^0_{1,02} + R^3_{1,32}   +  \frac{m}{2r^3} = 0 \].  \\       
It follows that the $4$ central second order CC of the list successively amounts to $R_{12}=0, R_{13}=0,R_{02}=0, R_{03}=0$, a result breaking the intrinsic/coordinate-free interpretation of the $10$ Einstein equations and the situation is even worst for the other components of the Ricci tensor. Indeed, $R_{01}$ and $R_{23}$ only depend on the vanishing of $R_{02,12}, R_{03,13}$ and $R_{02,03}, R_{12,13}$ among the bottom CC of the list, while the diagonal terms
$R_{00}, R_{11}, R_{22}, R_{33}$ only depend, {\it as we just saw}, on the $6$ non zero components of the Riemann tensor (!). We have thus obtained the totally unusual partition $10= 4 + 4 + 2$ along the successive blocks of the former list with:  \\
\[   \{ R_{ij}\}=\{ R_{00},R_{11},R_{22},R_{33}\}+ \{ R_{12},R_{13},R_{02},R_{03}\} + \{R_{01},R_{23}\}  \]  

Finally, we have to notice that $R_{01,23}=0,R_{02,31}=0\Rightarrow R_{03,12}=0$ from the identity:  \\
\[    R \in ker(\delta) \,\,\, \Rightarrow \,\,\, R_{01,23} + R_{02,31} + R_{03,12}=0  \]
and there is no way to have two identical indices in the first jets appearing through the (formal) Lie derivative just described. As for the third order CC, setting ${\xi}_{1,1}=\frac{1}{2}{\Omega}_{11} - \frac{A'}{2A}{\xi}_1\in j_2(\Omega)$, we have at least the first prolongations of the previous second order CC to which we have to add the new generating (where the first is the identity $0=0$):     \\
\[ d_0{\xi}_1-{\xi}_{1,0}=0,  d_1{\xi}_1 - {\xi}_{1,1}=0,d_2{\xi}_1-{\xi}_{1,2}=0, d_3{\xi}_1-{\xi}_{1,3}=0, d_2{\xi}_{0,3}-d_3{\xi}_{0,2}=0 \]
provided by the Spencer operator, leading to the crossed terms $d_i{\xi}_{1,0} - d_0{\xi}_{1,i}=0, \forall i=1,2,3$ and $d_i{\xi}_{1,j} - d_j{\xi}_{1,i}=0, \forall i,j=1,2,3$ because the Spencer operator is not FI. We shall discover later on that we have indeed only $4$ generating third order CC. Of course, in each of the preceding situations, we obtain the corresponding CC by replacing the jets by their expressions in terms of $j_2(\Omega)$ or 
$j_3(\Omega)$.\\

Such striking results are brought by the formal Lie derivative of the Weyl tensor because the Ricci tensor vanishes by assumption and we have the splitting $Riemann\simeq Ricci \oplus Weyl$ according to the {\it fundamental diagram II} that we discovered as early as in 1988 ([22])  but is still not acknowledged though it can be found in ([23],[27],[30-31]). In particular, as the $Ricci$ part is vanishing by assumption, we may identify the $Riemann$ part with the $Weyl$ splitting part as tensors ([31], Th 4.8 and [33]) and it is possible to prove (using a tedious direct computation or computer algebra) that the only $6$ non-zero components are the ones just used in the above Remark. It is important to notice that this result, bringing a strong condition on the zero jets because of the Lie derivative of the Weyl tensor and thus on the first jets, involves indeed the first derivative of the Weyl tensor because we have a term in $(A'')'$. However, as we are dealing with sections, ${\xi}_1=0$ implies ${\xi}_{1,1}=0$ and we also have ${\xi}_{0,0}=0, {\xi}_{1,2}=0,{\xi}_{1,3}=0$ but {\it not} ${\xi}_{1,0}=0$, these later condition being only brought by one additional prolongation and we have the strict inclusions  $R^{(3)}_1\subset R^{(2)}_1 \subset R^{(1)}_1=R_1$. Hence, it remains to determine the dimensions of these subsystems, exactly again like in the Macaulay or Janet examples. Knowing that $dim(R_1)=dim(R_2)=10$, $ dim(R_3) = 5$, $dim(R_4)= 4$, we have thus obtained the $15$ equations defining $R_3$ with $dim(R_3)=20-15=5$ and the $16$ equations defining $R_4$ with $dim(R_4)=20-16=4$, namely:  \\
\[  \begin{array}{l}
{\xi}_{3,3}+ sin(\theta)cos(\theta) {\xi}_2 =0 \\
{\xi}_{2,3} +{\xi}_{3,2}-2cot(\theta)\, {\xi}_3  =0  \\
{\xi}_{1,3}  =0  \\
{\xi}_{0,3} =0  \\
{\xi}_{2,2} =0 \\
{\xi}_{1,2} =0  \\
{\xi}_{0,2}  =0  \\
{\xi}_{3,1} - \frac{2}{r}\,{\xi}_3 =0  \\
{\xi}_{2,1} - \frac{2}{r} \,{\xi}_2  =0  \\
{\xi}_{1,1} =0  \\
{\xi}_{0,1}  - \frac{A'}{A}\,{\xi}_0=0 \\
{\xi}_{3,0}=0  \\
{\xi}_{2,0}=0  \\
{\xi}_{1,0}=0  \\
{\xi}_{0,0} =0  \\
{\xi}_1=0
\end{array}   \]
Setting now in a non-intrinsic way ({\it care}), ${\xi}_0=A{\xi}^0, {\xi}_1= - \frac{1}{A}{\xi}^1, {\xi}_2= - r^2{\xi}^2, {\xi}_3= - r^2{\xi}^3$, we may even simplify these equations and get a system {\it not depending on} $A$ {\it anymore}: \\
\[  \left\{\begin{array}{l}
{\xi}^3_3+ sin(\theta)cos(\theta) {\xi}^2 =0 \\
{\xi}^2_3 +{\xi}^3_2 - 2cot(\theta)\, {\xi}^3  =0  \\
{\xi}^1_3  =0  \\
{\xi}^0_3 =0  \\
{\xi}^3_1  =0  \\
{\xi}^2_1   =0  \\
{\xi}^1_1 =0  \\
{\xi}^0_1  =0 \\
{\xi}^3_0  =0  \\
{\xi}^2_0=0  \\
{\xi}^1_0=0  \\
{\xi}^0_0 =0  \\
{\xi}^2_2 =0 \\
{\xi}^1_2 =0  \\
{\xi}^0_2  =0  \\
{\xi}^1=0
\end{array}   \right.
\fbox{ $ \begin{array}{cccc}
2& 0 & 1 & 3  \\
2 & 0 &1 & 3  \\
2 & 0 & 1 & 3 \\
2 & 0 & 1 & 3 \\
2 & 0 & 1 & \bullet  \\
2 & 0 & 1 & \bullet  \\
2& 0 & 1 & \bullet  \\
2 & 0 & 1 & \bullet  \\
2 & 0 & \bullet & \bullet  \\
2 & 0  & \bullet & \bullet  \\
2 & 0 & \bullet & \bullet   \\
2 & 0 & \bullet & \bullet  \\
2 & \bullet & \bullet  & \times  \\
2 & \bullet & \bullet & \bullet  \\
2 & \bullet & \bullet & \bullet  \\
\bullet & \bullet & \bullet & \bullet
\end{array}  $ }   \]
and we have replaced by "$\times$" the only dot that cannot provide vanishing crossed derivatives and thus involution, being left with $24$ first order CC.
It is easy to check that $R^{(3)}_1$, having minimum dimension equal to $4$, is formally integrable, though not involutive as it is finite type, and to exhibit $4$ solutions linearly independent over the constants. Indeed, we must have ${\xi}^0=c$ where $c$ is a constant and we may drop the time variable not appearing elsewhere while using the equation ${\xi}^1=0$. It follows that ${\xi}^2=f(\theta,\phi), {\xi}^3=g(\theta, \phi)$ while $f,g$ are solutions of the first, second and fifth equations of Killing type wih a general solution depending on $3$ constants, a result leading to an elementary probem of $2$-dimensional elasticity left to the reader as an exercise. The system $R^{(3)}_1$ is formally integrable while the system $R^{(2)}_2$ is involutive. Having in mind the PP procedure, it follows that the CC could be of order $2,3$ {\it and even} $4$ along the following commutative and exact diagrams:\\
  \[  \begin{array}{rccccccccccl}
 &&&& 0 &&0&& &  & \\
 &&&& \downarrow && \downarrow &&   \\
  && 0 & \longrightarrow &S_4T^*\otimes T&\longrightarrow&S_3T^*\otimes F_0 &\longrightarrow & h_3&\longrightarrow  0  \\
   &&\downarrow && \downarrow & & \downarrow & \searrow  & \downarrow & &   & \\
0& \longrightarrow &R_4&\longrightarrow &J_4(T)& \longrightarrow &J_3(F_0) &\longrightarrow & Q_3 &   \longrightarrow 0 \\
    & & \downarrow && \hspace{5mm}\downarrow {\pi}^4_3 & &\hspace{5mm} \downarrow {\pi}^3_2 & & \downarrow  &  \\
   0 & \longrightarrow & R_3& \longrightarrow & J_3(T) & \longrightarrow &J_2(F_0)& \longrightarrow &Q_2& \longrightarrow  0 \\
   && && \downarrow & & \downarrow & & \downarrow &      \\
   &&  && 0 && 0 && 0  &  
   \end{array}     \]
 \[  \begin{array}{rccccccccccl}
 &&&& 0 &&0&& &  & \\
 &&&& \downarrow && \downarrow &&  \\
  && 0 & \longrightarrow &140 &\longrightarrow& 200 &\longrightarrow & 60 &\longrightarrow  0  \\
   &&\downarrow && \downarrow & & \downarrow &\searrow & \downarrow & &   & \\
0& \longrightarrow &4&\longrightarrow & 280 & \longrightarrow & 350 &\longrightarrow & 74 &   \longrightarrow 0 \\
    & & \downarrow && \hspace{5mm}\downarrow {\pi}^4_3 & &\hspace{5mm} \downarrow {\pi}^3_2 & & \downarrow  &  \\
   0 & \longrightarrow & 5& \longrightarrow & 140 & \longrightarrow & 150 & \longrightarrow &15 & \longrightarrow  0 \\
   &&&& \downarrow & & \downarrow & & \downarrow &      \\
   && && 0 && 0 && 0  &  
   \end{array}     \]
 
   \[  \begin{array}{rccccccccccl}
 &&&& 0 &&0&&0 &  & \\
 &&&& \downarrow && \downarrow && \downarrow  \\
  && 0 & \longrightarrow &S_5T^*\otimes T&\longrightarrow&S_4T^*\otimes F_0 &\longrightarrow & h_4&\longrightarrow  0  \\
   &&\downarrow && \downarrow & & \downarrow & \searrow  & \downarrow & &   & \\
0& \longrightarrow &R_5  &\longrightarrow &J_5(T)& \longrightarrow &J_4(F_0) &\longrightarrow & Q_4 &   \longrightarrow 0 \\
    & & \downarrow && \hspace{5mm}\downarrow {\pi}^4_3 & &\hspace{5mm} \downarrow {\pi}^3_2 & & \downarrow  &  \\
   0 & \longrightarrow & R_4 & \longrightarrow & J_4(T) & \longrightarrow &J_3(F_0)& \longrightarrow &Q_3& \longrightarrow  0 \\
   &&\downarrow && \downarrow & & \downarrow & & \downarrow &      \\
   && 0 && 0 && 0 && 0  &  
   \end{array}     \]
 \[  \begin{array}{rccccccccccl}
 &&&& 0 &&0&&0 &  & \\
 &&&& \downarrow && \downarrow && \downarrow  \\
  && 0 & \longrightarrow &224&\longrightarrow& 350 &\longrightarrow & 126 &\longrightarrow  0  \\
   &&\downarrow && \downarrow & & \downarrow &\searrow & \downarrow & &   & \\
0& \longrightarrow &4&\longrightarrow & 504 & \longrightarrow & 700 &\longrightarrow & 200 &   \longrightarrow 0 \\
    & & \downarrow && \hspace{5mm}\downarrow {\pi}^4_3 & &\hspace{5mm} \downarrow {\pi}^3_2 & & \downarrow  &  \\
   0 & \longrightarrow & 4& \longrightarrow & 280& \longrightarrow & 350 & \longrightarrow &74 & \longrightarrow  0 \\
   &&\downarrow && \downarrow & & \downarrow & & \downarrow &      \\
   && 0 && 0 && 0 && 0  &  
   \end{array}     \] 
   
As we have at least one third order CC, we cannot choose $F_1=Q_2$. If we choose $F_1=Q_4$, we have indeed $dim(R_{5+r})=dim(R_5)=dim(R_4)=4,\forall r\geq 0$ and the short exact sequences: \\
\[  0 \rightarrow R_{5+r} \rightarrow J_{5+r}(T) \rightarrow J_{4+r}(F_0) \rightarrow J_r(F_1)    \]
by applying the Spencer $\delta$-map inductively to the symbol sequences:  \\
\[ 0 \rightarrow S_{5+r}T^*\otimes T \rightarrow S_{4+r}T^*\otimes F_0 \rightarrow S_rT^*\otimes F_1  \]
and chasing as usual along the south-west to north-east diagonal. After tedious computations, we obtain in this way the formally exact differential sequence:   \\
\[    0 \rightarrow \Theta \rightarrow 4 \underset 1{\rightarrow} 10 \underset 4{\rightarrow} 200  \underset 1{\rightarrow} 576 \underset 1{\rightarrow} 664 
\underset 1{\rightarrow} 356 \underset 1{\rightarrow} 74  \rightarrow 0  \]
with Euler-Poincar\'{e} characteristic $rk_D(M)=4 - 10 + 200 - 576 + 664 - 356 + 74 = 0 $.  \\

It remains to prove that, exactly like in the case of the Macaulay Example 2A.9 that needed $2$ prolongations and had only second order CC, here we need $3$ prolongations and could obtain only third order CC by choosing $F_1=Q_3$. 
For such a purpose and {\it in this particular situation only}, we may introduce the vertical first Spencer sequences in the commutative diagram below where $d$ is the Spencer operator of Proposition $2C.3$, exactly like in ([20], p 190) or ([21], p 688). As it is well known that the two central columns are localy exact, we deduce from a chase that the upper row is locally exact at $F_0$, and thus ${\cal{D}}_1 $ generates the CC of ${\cal{D}}$, if and only if the left vertical Spencer sequence is locally exact at $T^*\otimes R_3$.  \\
\[  \begin{array}{rcccccccl}
      &  0\,  & &   0 \,  &  &   0 \, &   &  &    \\
   &  \downarrow \, & & \downarrow , &   &  \downarrow \, &  & & \\
0 \longrightarrow  &  \Theta & \longrightarrow & T  &  \stackrel{{\cal{D}}}{\longrightarrow} & F_0  & \stackrel{{\cal{D}}_1}{\longrightarrow}   &  F_1 &  \\
   & \, \,\,\, \downarrow j_4  &  & \, \,\,\, \downarrow j_4  &  &   \,\,\,\, \downarrow j_3  & \searrow &  \parallel &  \\
   0 \longrightarrow &  R_4  & \longrightarrow & J_4(T) & \longrightarrow & J_3(F_0) & \longrightarrow  & F_1  & \longrightarrow  0 \\
     &  \,\,\,\, \downarrow d &  &  \, \,\,\, \downarrow d  &  &  \,\,\,\, \downarrow d  &  &  &    \\
     0  \longrightarrow &  T^*\otimes R_3&  \longrightarrow  &  T^*\otimes J_3(T) & \longrightarrow  & 
     T^*\otimes J_2(F_0) &  \\
     & \, \,\,\,  \downarrow d  &  & \, \,\,\,  \downarrow d &  &  &  &  \\
     0  \longrightarrow & {\wedge}^2T^*\otimes R_2 & \longrightarrow  &{\wedge}^2T^*\otimes  J_2(T) &  &  &
\end{array}  \]

Now, we have the commutative diagram:  \\

\[  \begin{array}{rccccccc}
&  &  &  0  &  &  0  &  &  0    \\
 &    &  & \downarrow &  &  \downarrow &  &  \downarrow  \\
 0 \longrightarrow &  \Theta  &  \stackrel{j_5}{\longrightarrow} & R_5 & \stackrel{d}{\longrightarrow} & T^*\otimes R_4 &  \stackrel{d}{\longrightarrow } & {\wedge}^2T^*\otimes R_3  \\
   &  \parallel  &  &  \downarrow &  &  \downarrow &  &  \downarrow  \\
 0 \longrightarrow &  \Theta  &  \stackrel{j_4}{\longrightarrow} & R_4 & \stackrel{d}{\longrightarrow} & T^*\otimes R_3 &  \stackrel{d}{\longrightarrow } & {\wedge}^2T^*\otimes R_2  \\
  &  &  &  \downarrow &  &  \downarrow &  &  \downarrow    \\
  &  &  &  0  &  \longrightarrow  & T^*\otimes (R_3/R^{(1)}_3) & \stackrel{d}{\longrightarrow} & {\wedge}^2T^*\otimes (R_2/R^{(1)}_2)  \\
  &  &  &  &  & \downarrow & & \downarrow   \\
  &  &  &  &  &    0  &  &  0  
\end{array}  \]

As the second Spencer sequence {\it in this specific example} is locally isomorphic to the tensor product of the Poincar\'{e} sequence for $n=4$ by a Lie algebra of dimension $4$ ($1$ time translation+ $3$ space rotations), it is therefore locally exact and the upper row of the last diagram is thus locally exact at $T^*\otimes R_4$. It follows from a chase that the central row is locally exact at $T^*\otimes R_3$ as we wished, if ({\it but not only if} !) the bottom row is exact. Using the fact that $dim(R_3/R^{(1)}_3)=1$ and $dim(R_2/R^{(1)}_2)=dim(R^{(1)}_1/R^{(2)}_1)= 10-5=5$, we may choose a representative in $T^*\otimes R_3$ in such a way that all the zero jets are zero and all the 
first jets vanish but $v_{1,0}=-v_{0,1}$. Coming back to the standard notation $v^k_{\mu,I}$ of the formal theory without specifying the residue by a bar for simplicity while applying the $\delta$ map, we get successively:  \\
\[  v^1_{,0i}\equiv v^1_{0,i} -v^1_{i,0}=0  \Rightarrow v^1_{0,i}=0, \forall i=1,2,3  \]
and the lower $d$ is not injective. If we suppose nevertheless that the central row is locally exact, going on this way inductively, we should obtain by tedious computations the {\it simplest, shortest, formally and locally exact} differential sequence:  \\
\[  0 \rightarrow \Theta \rightarrow 4 \underset 1{\rightarrow}10 \underset 3{\rightarrow} 74 \underset 1{\rightarrow} 170 \underset 1{\rightarrow} 164 \underset 1{\rightarrow} 76 \underset 1{\rightarrow} 14 \rightarrow 0    \]
with Euler-Poincar\'{e} characteristic  $ rk_D(M)=4 - 10 +74 - 170 + 164 - 76 + 14=0$. We wish good luck to anybody trying to obtain such a sequence by means of computer algebra as certain matrices are up to $9324 \times 11900$ and do not therefore believe that such a sequences could be useful in physics. We finally notice that, on the contrary, this method can also be {\it safely} used whith the previous choice $F_1=Q_4$ and $R_5$ in place of $R_4$ because, {\it in this case}, the Spencer sequence $R_5 \stackrel{d}{\longrightarrow} T^*\otimes R_4 \stackrel{d}{\longrightarrow} {\wedge}^2T^*\otimes R_3$ is locally exact and we should obtain the sequence already exhibited.  \\

However, this fact is of no importance compared to the following final comments that we shall provide below and will be explained later on in a much simpler and direct way.  \\

First of all, denoting by $R'_2=R^{(2)}_2\subset R_2\subset J_2(T)$ with $dim(R'_2)=4$ the involutive system provided by the PP procedure, we are in position to construct the corresponding canonical/involutive (lower) Janet and (upper) Spencer sequences along the following {\it fundamental diagram I} already constructed in many books and papers (In particular, we advise the curious reader to look at the striking Example $3.14$ described in [29], p 119 and showing the importance of involution) and already presented in the last subsection $A$. In this diagram, {\it not depending any longer on} $m$, we have now $C_r={\wedge}^rT^*\otimes R'_2$ and provide the fiber dimensions below:  \\

 \[   \begin{array}{rcccccccccccccl}
 &&&&& 0 &&0&&0&& 0 &  &0&  \\
 &&&&& \downarrow && \downarrow && \downarrow & & \downarrow &   & \downarrow &  \\
  & 0& \rightarrow& \Theta &\stackrel{j_2}{\rightarrow}&4 &\stackrel{D_1}{\rightarrow}& 16 &\stackrel{D_2}{\rightarrow} & 24&\stackrel{D_3}{\rightarrow}&16 &\stackrel{D_4}{\rightarrow}& 4 &\rightarrow 0 \\
  &&&&& \downarrow & & \downarrow & & \downarrow & & \downarrow &&\downarrow &     \\
   & 0 & \rightarrow & 4 & \stackrel{j_2}{\rightarrow} & 60 & \stackrel{D_1}{\rightarrow} & 160 &\stackrel{D_2}{\rightarrow} & 180 &\stackrel{D_3}{\rightarrow} & 96 &\stackrel{D_4}{\rightarrow} &20&   \rightarrow 0 \\
   & & & \parallel && \downarrow  & & \downarrow  & & \downarrow  & & \downarrow & & \downarrow  & \\
   0 \rightarrow & \Theta &\rightarrow & 4 & \stackrel{\cal{D}}{\rightarrow} & 56 & \stackrel{{\cal{D}}_1}{\rightarrow} & 144 & \stackrel{{\cal{D}}_2}{\rightarrow} & 156 & \stackrel{{\cal{D}}_3}{\rightarrow} & 80 &\stackrel{{\cal{D}}_4}{\rightarrow} & 16 & \rightarrow  0 \\
   &&&&& \downarrow & & \downarrow & & \downarrow & & \downarrow &  &\downarrow &   \\
   &&&&& 0 && 0 && 0 &&0 &&0 &  
   \end{array}     \]
 We notice the vanishing of the Euler-Poincar\'{e} characteristics:  \\  
\[   4 - 16 +24 - 16 + 4=0, \hspace{2mm}  4 - 60 +160 - 180 + 96 - 20 =0, \hspace{2mm} 4 - 56 + 144 - 156 + 80 - 16 = 0   \]

We shall now use the above result in order to study directly the first order formally integrable system $R'_1=R^{(3)}_1\subset J_1(T)$ while replacing $F_0=J_1(T)/R_1=S_2T^*$ with $dim(F_0)=10$ by $F'_0=J_1(T)/R'_1$ with $dim(F'_0)=16$. We already know that $g'_2=g^{(2)}_2=g_2=0$ but we have to check whether $g'_1$ is $2$-acyclic or not before deciding about the CC of the corresponding operator ${\cal{D}}':T \rightarrow F'_0$ while using Theorem $2A.6$. For this, let us consider the following commutative diagram with exact rows and exact columns but the left or right ones:  \\

 \[  \begin{array}{rcccccccccccl}
 &&&& 0 &&0&&&  & &  \\
 &&&& \downarrow && \downarrow && & &  & \\
  && 0 & \rightarrow &S_3T^*\otimes T&\rightarrow&S_2T^*\otimes F'_0 &\rightarrow & 80&
   \rightarrow  & 0  \\
   &&&& \hspace{3mm} \downarrow \delta & & \hspace{3mm} \downarrow \delta & & \downarrow & &  &  & \\
&  &0&\rightarrow &T^*\otimes S_2T^*\otimes T & \rightarrow &T^*\otimes T^* \otimes F'_0 &\rightarrow & 96 &   \rightarrow  &   0 \\
    & & && \hspace{3mm}\downarrow \delta  & &\hspace{3mm} \downarrow \delta& & \downarrow  &  &  \\
   0 & \rightarrow & {\wedge}^2T^*\otimes g'_1& \rightarrow & {\wedge}^2T^*\otimes T^* \otimes T & \rightarrow & {\wedge}^2T^*\otimes F'_0  & \rightarrow & 6 &  \rightarrow &  0  \\
   && \hspace{3mm} \downarrow \delta && \hspace{3mm} \downarrow \delta & & \downarrow & & \downarrow  &  &  &  \\
 0   & \rightarrow & {\wedge}^3T^*\otimes T & = & {\wedge}^3T^*\otimes T & \rightarrow & 0  & &  0  &&  & \\
       &   &  &  & \downarrow  &  &  &  &  &  & &  \\
   &&&& 0 && && &  &
   \end{array}     \]

\noindent
{\bf LEMMA 3B.2}: $g'_1$ is {\it not} $2$-acyclic and the CC of ${\cal{D}}'$ are of order $2$, thanks to Theorem $2A.6$.  \\
\noindent
\begin{proof}
We have $dim(g'_1)=1$ and the only non-zero parametric jet of the symbol is ${v}^3_2 = - {v}^2_3$. Therefore, we just need to prove that the map $\delta$ in the left column is not injective and thus also not surjective. We have $16$ equations like $ v^0_{,123}=v^0_{1,23} + v^0_{2,31} + v^0_{3,12}=0$ providing $0=0$ or like $v^2_{,123}\equiv v^2_{1,23} + v^2_{2,31} + v^2_{3,12}=0$ and thus $v^2_{3,12}=0$. However, as $v^2_{3,23}$ cannot appear in the formulas, we obtain $dim(H^2(g'_1))=1$ with the long exact ker/coker sequence $0 \rightarrow 1 \rightarrow 6 \stackrel{\delta}{\rightarrow}  16 \rightarrow 11 \rightarrow 0$, that is the $\delta$-sequence $0 \rightarrow 6 \stackrel{\delta}{\rightarrow} 16 \rightarrow 0$ has a cohomology of dimension $1$ at $6$ and a cohomology of dimension $11$ at $16$. This result is proving that the upper map $80 \rightarrow 96$ in the right column is not injective with a kernel of dimension $1$ and an image of dimension $79< 96$ while the lower map $96 \rightarrow 6$ in the right column is indeed surjective with a kernel of dimension $90<96$, that is we have a cohomology of dimension $90-79=11$ at $96$, in a coherent way with the snake chase. \\
\end{proof}

Then, $F'_1$ with $dim(F'_1)=104$ is defined by the commutative and exact diagram:  \\

   \[  \begin{array}{rccccccccccl}
 &&&& 0 &&0&&0 &  & \\
 &&&& \downarrow && \downarrow && \downarrow  \\
  && 0 & \longrightarrow &S_3T^*\otimes T&\longrightarrow&S_2T^*\otimes F'_0 &\longrightarrow & h'_2&\longrightarrow  0  \\
   &&\downarrow && \downarrow & & \downarrow & \searrow  & \downarrow & &   & \\
0& \longrightarrow &R'_3  &\longrightarrow &J_3(T)& \longrightarrow &J_2(F'_0) &\longrightarrow & F'_1 &   \longrightarrow 0 \\
    & & \downarrow && \hspace{5mm}\downarrow {\pi}^3_2 & &\hspace{5mm} \downarrow {\pi}^2_1& & \downarrow  &  \\
   0 & \longrightarrow & R'_2 & \longrightarrow & J_2(T) & \longrightarrow &J_1(F'_0)& \longrightarrow &Q'_1& \longrightarrow  0 \\
   &&\downarrow && \downarrow & & \downarrow & & \downarrow &      \\
   && 0 && 0 && 0 && 0  &  
   \end{array}     \]
It follows that, contrary to the Minkowski example where we had $F_1=H^2(g_1)$, in the Schwarzschild example we have indeed $F'_1\neq H^2(g'_1)$ by counting the dimensions, even though we have the strict inclusions $R'_1\subset R_1$ with $4<10$ and $g'_1 \subset g_1$ with $1< 6$. However, though the strict inclusion $R'_1\subset R_1$ induces an epimorphism $F'_0 \rightarrow F_0$ and thus an epimorphism $h'_2 \rightarrow h_2=F_1$, there is no way to link $F'_1$ with $F_1$ because of the additional $24$ CC of order $1$ already described. In particular, setting $\{{\xi}_1=U, {\xi}_{1,2}=V_2, {\xi}_{1,3}=V_3, {\xi}_{0,2}=W_2, {\xi}_{0,3}=W_3\}\in j_2(\Omega)$, we have successively:  \\
\[  {\xi}_{0,0}= \frac{1}{2}{\Omega}_{00} + \frac{AA'}{2}U, \,\,\,{\xi}_{0,1} - \frac{A'}{A}{\xi}_0={\Omega}_{01}- \fbox {$d_0U$}\in j_3(\Omega), \,\,\,{\xi}_{0,2}= W_2, \,\,\,{\xi}_{0,3}=W_3 \]
\[  {\xi}_{1,0} = \fbox{$d_0U$}\in j_3(\Omega), \,\,\, {\xi}_{1,1} =\frac{1}{2}{\Omega}_{11}-\frac{A'}{2A}U, \,\,\, {\xi}_{1,2}=V_2, \,\,\, {\xi}_{1,3}=V_3 \]
\[ {\xi}_{2,0}= {\Omega}_{02} -W_2, \,\,\, {\xi}_{2,1} - \frac{2}{r}{\xi}_2= {\Omega}_{12} - V_2, \,\,\, {\xi}_{2,2}= \frac{1}{2}{\Omega}_{22} - rAU \]
\[ {\xi}_{3,0}= {\Omega}_{03} - W_3, \,\,\, {\xi}_{3,1} - \frac{2}{r}{\xi}_3= {\Omega}_{13} -V_3 , \,\,\, {\xi}_{3,2}+{\xi}_{2,3} -
2cot(\theta){\xi}_3={\Omega}_{23}, \]  
\[{\xi}_{3,3}+ sin(\theta)cos(\theta){\xi}_2= \frac{1}{2}{\Omega}_{33} - rAsin^2(\theta) U  \]
First of all, as already said, we have $15$ second order CC in $j_2(\Omega)$ and their formal derivatives. Then, we have identities to zero like $d_0{\xi}^1-{\xi}^1_0=0$ but we have also {\it surely} the three third order CC like $d_1{\xi}^1- {\xi}^1_1=0, d_2{\xi}^1-{\xi}^1_2, d_3{\xi}^1-{\xi}^1_3$, then {\it  perhaps} the other third order CC $d_2{\xi}^0_3 - d_3{\xi}^0_2=0$ and {\it perhaps} even fourth order CC like $d_0{\xi}^0_1 - d_1{\xi}^0_0=0$ which is containing the leading term $d_{00}U$ after substitution. However, we have successively: \\ 
\[  \begin{array}{rcl}
    \frac{2r}{3m\,\,sin^2(\theta)}R_{23,30} + \frac{2}{3}{\Omega}_{02} & = \,\,{\xi}_{0,2}\,\, = & - \frac{2r^3A}{3m}R_{01,12}+\frac{1}{3}{\Omega}_{02}  \\                                                            
  \frac{2r}{3m}R_{23,02} +\frac{2}{3}{\Omega}_{03} & = \,\, {\xi}_{0,3} \,\, = &  - \frac{2r^3A}{3m}R_{01,13} +\frac{1}{3}{\Omega}_{03} 
\end{array}  \]
now, denoting by $\nabla$ the covariant derivative, we get by linearization with symbol " $lin $ ":  \\
\[  {\omega}_{kr}{\gamma}^k_{ij}=\frac{1}{2} ({\partial}_i{\omega}_{rj} + {\partial}_j{\omega}_{ir} - {Ê\partial}_r{\omega}_{ij})  \stackrel{lin}{\rightarrow } {\omega}_{kr} {\Gamma}^k_{ij} + {\gamma}^k_{ij}{\Omega}_{kr}= \frac{1}{2} (d_i{\Omega}_{rj} + d_j{\Omega}_{ir} - d_r{\Omega}_{ij})  \]
\[\nabla \rho=\partial \rho - \Sigma \gamma \rho \stackrel{lin}{\rightarrow} d R - \Sigma (\gamma R + \rho \Gamma)=\nabla R -\Sigma \rho \Gamma \]
\[ {\nabla}_3{\rho}_{01,12}= {\partial}_3{\rho}_{01,12} \stackrel{lin}{\rightarrow }d_3R_{01,12}-{\gamma}^3_{23}R_{01,13} - {\rho}_{21,12}{\Gamma}^2_{03}\]
Linearizing the Bianchi identity ${\nabla}_1{\rho}_{01,23} + {\nabla}_2{\rho}_{01,31} + {\nabla}_3{\rho}_{01,12}=0$, we get:  \\
\[  d_1R_{01,23} + d_2R_{01,31}+d_3R_{01,12} - {\gamma}^3_{23} (R_{01,31} + R_{01,13}) - {\rho}_{13,13}{\Gamma}^3_{02} - 
{\rho}_{12,12}{\Gamma}^2_{03}=0  \]
and thus:  \\
\[   R_{01,23}=0 \Rightarrow d_2R_{01, 13} - d_3R_{01, 12}- \frac{m}{2rA}{\Gamma}^2_{03} + \frac{msin^2(\theta)}{2rA}{\Gamma}^3_{02}=0 \Leftrightarrow  d_2{\xi}_{0,3} - d_3{\xi}_{0,2}=0  \]
a result showing that the third order CC $d_2W_3-d_3W_2=0$ is not a generating one because it is just a differential consequence of the second order CC 
$ R_{01,23}=0$.  \\

We also have: \\
\[  R_{01,02}= - \frac{3mA}{2r^3} {\xi}_{1,2} + \frac{mA}{2r^3}{\Omega}_{12}, \,\,\, R_{01,03}= - \frac{3mA}{2r^3} {\xi}_{1,3} +
 \frac{mA}{2r^3}{\Omega}_{13}  \]
\[  d_0R_{01,23} + d_2R_{01,30} +d_3R_{01,02} + (j_1(\Omega))=0          \]
a result showing that the third order CC $d_2V_3-d_3V_2=0$ is not a generating one because it is just a differential consequence of the second order CC $R_{01,23}=0$.  \\

Similarly, working modulo $j_1(\Omega)$, we should obtain by crossed derivatives the third order CC:  \\
\[   d_0{\xi}_{2,2} - d_2{\xi}_{2,0}= d_2 W_2 - rAd_0U + (j_1(\Omega)) =0 \]
while we have:  \\
\[ R_{12,12}= \frac{3m}{2r^2}{\xi}_1, \,\,\, R_{12,20}=0, \,\,\, R_{12,01}= - \frac{3m}{2r^3A}{\xi}_{0,2} \]
\[  d_0R_{12,12} + d_1R_{12,20} + d_2R_{12,01} + (j_1(\Omega))=0  \]
a result showing that the preceding third order CC is not a generating one because it is just a differential consequence of the second order CC 
$R_{12,02}=0$.  \\

It is even more difficult to check that the vanishing of the crossed derivatives:  \\
\[ d_0{\xi}_{3,1} -d_1{\xi}_{3,0}=0 \,\, \Leftrightarrow \,\, d_1W_3-d_0V_3-\frac{2}{r} W_3+(d_0{\Omega}_{13} - d_1{\Omega}_{03}+ 
\frac{2}{r}{\Omega}_{03})=0  \]
which can be checked directly, is in fact coming from the Bianchi identity:  \\
\[  d_0R_{23,12} + d_1R_{23,20} + d_2R_{23,01} \in j_1(\Omega)  \]
Indeed, we notice that: \\
\[   \begin{array}{rcrcl} 
R^2_{3,12} & =  &  - \frac{1}{r^2}R_{23,12} & = & \frac{3m}{2r^3}{\xi}_{3,1} - \frac{3m}{r^4}{\xi}_ 3 - \frac{m}{2r^3}{\Omega}_{13}   \\
R^2_{3,20} & =  &  - \frac{1}{r^2}R_{23,20} & =  & - \frac{3m}{2r^3}{\xi}_{3,0} + \frac{m}{2r^3}{\Omega}_{03}   \\  
R^2_{3,01} & =  &   - \frac{1}{r^2}R_{23,01} & = & 0  \\
R^1_{3,01} & = & - A R_{13,01} & = & - \frac{3m}{2r^3}{\xi}_{3,0} + \frac{m}{r^3}{\Omega}_{03}
\end{array}   \]  
The corresponding covariant dervatives are:  \\    
\[ \begin{array}{rcl}
 {\nabla}_0R^2_{3,12}& = & d_0R^2_{3,12}  + {\gamma}^0_{01}R^2_{3,20} \\
{\nabla}_1R^2_{3,20} & = & d_1R^2_{3,20} + {\gamma}^2_{r1}R^r_{3,20}- {\gamma}^r_{13}R^2_{r,20} - {\gamma}^r_{12}R^2_{3,r0}- 
{\gamma}^r_{01}R^2_{3,2r} \\
{\nabla}_2R^2_{3,01} & = & d_2R^2_{3,01} + {\gamma}^2_{12}R^1_{3,01} - {\gamma}^r_{32}R^2_{r, 01} - {\gamma}^r_{02}R^2_{3,r1} -{\gamma}^r_{12}R^2_{3,0r} 
\end{array}   \]
Surprisingly, many terms vanish individually while groups of other terms vanish collectively and it follows that {\it this} cyclic sum of covariant derivatives is equal to:  \\
\[  d_0R^2_{3,12}+d_1R^2_{3,20}+\frac{1}{r}R^1_{3,01} - cot(\theta)R^2_{3,01} =  \frac{3m}{2r^3}(d_1{\xi}_{3,0} - d_0{\xi}_{3,1}) - \frac{m}{2r^3}(d_0{\Omega}_{13} - d_1{\Omega}_{03}+ \frac{1}{r}{\Omega}_{03})   \]  
because $ R^2_{3,01}=0$ but {\it this} cyclic sum of the covariant derivatives must be corrected by adding terms linear in $\Gamma$, namely:  \\ 
\[  -{\rho}^2_{1,12} {\Gamma}^1_{03} - {\rho}^2_{3,32}{\Gamma}^3_{01}-{\rho}^2_{0,20} {\Gamma}^0_{13} -{\rho}^2_{3,23}{\Gamma}^3_{01} =  {\rho}^2_{1,21} {\Gamma}^1_{03} -{\rho}^2_{0,20} {\Gamma}^0_{13} = - \frac{m}{2r^3A}{\Gamma}^1_{03} - \frac{mA}{2r^3}{\Gamma}^0_{13}   \]
Using the previous linearizing formulas, we obtain:   \\
\[  \begin{array}{rclcl}
{\omega}_{11}{\Gamma}^1_{03} + {\gamma}^r_{03}{\Omega}_{1r} & = & -\frac{1}{A}{\Gamma}^1_{03} & = &
\frac{1}{2}( d_0{\Omega}_{13} + d_3{\Omega}_{01} - d_1{\Omega}_{03})   \\
{\omega}_{00}{\Gamma}^0_{13}+ {\gamma}^3_{13}{\Omega}_{03}  & = &  \,\,\,\,\,\,  A {\Gamma}^0_{13} + \frac{1}{r}{\Omega}_{03}  &  =  &
\frac{1}{2}( d_1{\Omega}_{03} + d_3{\Omega}_{01} - d_0{\Omega}_{13})   
\end{array}   \]
After multiplication of each equation by $\frac{m}{2r^3}$ and substraction, the above additional term becomes:  \\
\[    \frac{m}{2r^3}(d_0{\Omega}_{13} - d_1{\Omega}_{03}+ \frac{1}{r}{\Omega}_{03})   \]
As the Bianchi identities provide identities to zero in a tensorial manner, that is in $H^3(g_1) \subset {\wedge}^3T^*\otimes g_1$, their linearizations do vanish too and provides the vanishing of the crossed derivatives under study, a result showing that the previous third order CC is not a generating one as it is just a differential consequence of the second order CC $R_{23,01}=0$. We let the reader prove in a similar way that the CC 
$d_0{\xi}_{2,1}-d_1{\xi}_{2,0}=0$ is a linear consequence of the Bianchi identity:\\ 
\[    d_0R_{23,13} + d_1R_{23,30} + d_3R_{23,01} \in j_1(\Omega)  \]
and the corresponding third order CC is not a generating one as it is just a differential consequence of the same second order CC $R_{23,01}=0$.  \\

In a quite different manner, we have:  \\
\[  d_1{\xi}_{0,0}- d_0{\xi}_{0,1}= d_{00}{\xi}_1+d_1(\frac{AA'}{2}{\xi}_1) -\frac{A'}{A}d_0{\xi}_0 + \frac{1}{2}d_1{\Omega}_{00}-d_0{\Omega}_{01}  \]
The reader may check alone that the right member vanishes identically, providing theefore what we called {\it identity to zero} in Example 2A.9. 
However, such a result can be simply obtained from the lack of formal integrability of the Spencer operator. Indeed, we just need to notice that:   \\
\[   d_1(d_0{\xi}_0 -{\xi}_{0,0}) - d_0( d_1{\xi}_0 - {\xi}_{0,1})=d_0{\xi}_{0,1} - d_1{\xi}_{0,0} =0 \]
The same argument also provides:   \\
\[   d_1(d_0{\xi}_1 -{\xi}_{1,0}) - d_0( d_1{\xi}_1 - {\xi}_{1,1})=d_0{\xi}_{1,1} - d_1{\xi}_{1,0} =0 \]
but the situation is now slightly different because $d_1{\xi}_1-{\xi}_{1,1}=d_1U - +\frac{A'}{2A}U - \frac{1}{2}{\Omega}_{11}=0$ is nothing else than one of the third order generating CC.  \\

Finally, as already noticed, the symbol $g'_1\subset g_1\subset T^*\otimes T$ is not involutive and even $2$-acyclic because otherwise there should only be first order CC for the right members defining the system $R'_1\subset J_1(T)$. As a byproduct, we have, at least on the symbol level, the second 
order CC:  \\
\[  d_{22}{\xi}_{3,3} +d_{33}{\xi}_{2,2} -d_{23}({\xi}_{3,2}+{\xi}_{2,3})=0   \]
and thus:  \\
\[ d_{22}(\frac{1}{2}{\Omega}_{33} -rAsin^2(\theta){\xi}_1 - sin(\theta)cos(\theta){\xi}_2) + d_{33}(\frac{1}{2}{\Omega}_{22} - rA{\xi}_1) - d_{23}({\Omega}_{23}+ 2cot(\theta){\xi}_3)=0  \]
containing surely $d_{22}{\xi}_1, \,\, d_{22}{\xi}_2, \,\, d_{33}{\xi}_1, \,\, d_{23}{\xi}_3$ and thus surely $d_2U,\,\,d_2V_2, d_3V_3$, producing therefore a third order CC that cannot be reduced by means of any Bianchi identity, that is we finally have $15$ generating second order CC and $4$ new generating third order CC, in a manner absolutely similar to that of Example $2.A9$.  \\  \\

In actual practice, {\it all the preceding computations have been finally used to reduce the Poincar\'e group to its subgroup made with only one time translation and three space rotations} !. On the contrary, we have proved during fourty years that one {\it must} increase the Poincar\'e group ($10$ parameters), first to the Weyl group ($11$ parameters by adding $1$ dilatation) and finally to the conformal group of space-time ($15$ parameters by adding $4$ elations) while only dealing with he Spencer sequence in order to increase the dimensions of the Spencer bundles and thus the number of corresponding potentials and fields. We wish therefore also good luck to the reader who should like to find back these results by using computer algebra !.\\ 

Now, in order to convince the reader that {\it only new methods} can allow to study the strange phenomenas happening in the constructions of CC (high order, sudden increase in the number of generators, ...), we shall turn over totally the previous approach and use a totally different point of view, having in mind that we already know the final formally integrable system $R_4\subset J_4(T)$ with $dim(R_4)=4$ or the equivalent involutive system $R^{(2)}_2\subset J_2(T)$. Of course, the same method could be used for other cases. For this, we use the known Killing vector ${\partial}_t$ and the zero order equation ${\xi}^1=0$ in order to restrict $T$ to a sub-vector bundle 
$E\subset T$ of dimension $2$ with section $({\xi}^2, {\xi}^3)$ in order to have a first order system with $3$ independent variables 
$(r,\theta, \phi)=(1,2,3)$ and $2$ unknowns, obtained by eliminating ${\xi}^1$ and ${\xi}^0$ as follows in order to get an equivalent system for ${\xi}^2,{\xi}^3$ for the variables $(1,2,3)$ with only $5$ equations:  \\
\[ \left\{ \begin{array}{l}
{\xi}^3_3+ sin(\theta)cos(\theta) {\xi}^2 =0 \\
{\xi}^2_3 +{\xi}^3_2 - 2cot(\theta)\, {\xi}^3  =0  \\
{\xi}^3_1  =0  \\
{\xi}^2_1   =0  \\
{\xi}^2_2 =0 \\
\end{array}   \right.
\fbox{ $ \begin{array}{cccc}
2 & 1 & 3  \\
2  &1 & 3  \\
2  & 1 & \bullet  \\
2  & 1 & \bullet  \\
2 &  \times & \bullet  
\end{array}  $ } \]

As a byproduct, we have the following commutative diagrams:  \\

 \[  \begin{array}{rccccccccccl}
 &&&& 0 &&0&&&  & \\
 &&&& \downarrow && \downarrow && & & \\
  && 0 & \rightarrow &S_3T^*\otimes E&\rightarrow&S_2T^*\otimes F_0 &\rightarrow & h_2 &
   \rightarrow  0  \\
   &&&& \hspace{3mm} \downarrow \delta & & \hspace{3mm} \downarrow \delta & & \downarrow & &    & \\
&  &0&\rightarrow &T^*\otimes S_2T^*\otimes E & \rightarrow &T^*\otimes T^* \otimes F_0 &\rightarrow & T^*\otimes h_1 &   \rightarrow 0 \\
    & & && \hspace{3mm}\downarrow \delta  & &\hspace{3mm} \downarrow \delta& & \downarrow  &  \\
   0 & \rightarrow & {\wedge}^2T^*\otimes g_1& \rightarrow & {\wedge}^2T^*\otimes T^* \otimes E & \rightarrow & {\wedge}^2T^*\otimes F_0  & \rightarrow & 0 &  \\
   && \hspace{3mm} \downarrow \delta && \hspace{3mm} \downarrow \delta & & \downarrow & & &   &   \\
 0   & \rightarrow & {\wedge}^3T^*\otimes E  & = & {\wedge}^3T^*\otimes E & \rightarrow & 0  & &&&   \\
       &   &  \downarrow &  & \downarrow  &  &  &  &  &  &  \\
   &&0&& 0 && && &  &
   \end{array}     \]

 \[  \begin{array}{rccccccccccl}
 &&&& 0 &&0&&&  & \\
 &&&& \downarrow && \downarrow && & & \\
  && 0 & \rightarrow &20&\rightarrow&30 &\rightarrow & 10 &
   \rightarrow  0  \\
   &&&& \hspace{3mm} \downarrow \delta & & \hspace{3mm} \downarrow \delta & & \downarrow & &    & \\
&  &0&\rightarrow &36 & \rightarrow &45 &\rightarrow & 9 &   \rightarrow 0 \\
    & & && \hspace{3mm}\downarrow \delta  & &\hspace{3mm} \downarrow \delta& & \downarrow  &  \\
   0 & \rightarrow & 3 & \rightarrow & 18 & \rightarrow & 15 & \rightarrow & 0 &  \\
   && \hspace{3mm} \downarrow \delta && \hspace{3mm} \downarrow \delta & & \downarrow & & &   &   \\
 0   & \rightarrow &  2  & = &  2  & \rightarrow & 0  & &&&   \\
       &   &  \downarrow &  & \downarrow  &  &  &  &  &  &  \\
   &&0&& 0 && && &  &
   \end{array}     \]

The next result points out the importance of the Spencer $\delta$-cohomology and will be justified later on in cartesian coordinates as it is intrinsic:  \\

\noindent 
{\bf LEMMA 3B.3}: The last symbol diagram is commutative and exact. In particular, the lower left map $\delta$ is surjective and thus  the upper right induced map $h_2 \rightarrow T^*\otimes Q_1$ is  also surjective while these two maps have isomorphic kernels.  \\

\begin{proof}
The $3$ components of ${\wedge}^2T^*\otimes g_1$ are $\{ v^3_{2, 12}, v^3_{2,13}, v^3_{2,23}\}$ and the kernel of the map $\delta$ is described by the two linear equations:  \\
\[  v^2_{1,23} +v^2_{2,31} + v^2_{3,12}=0, \hspace{5mm}  v^3_{1,23} +v^3_{2,31} + v^3_{3,12}=0 \]
that is to say by the two linearly independent equations:  \\
\[    v^3_{2,12} =0, \hspace{5mm} v^3_{2,13} =0 \]
Accordingly, in the left column we have:  \\
\[    dim(H^2(g_1))=dim(Z^2(g_1))=dim(ker(\delta))=1\]   
An unusual snake-type diagonal chase left to the reader as an exercise proves that the induced map $h_2\rightarrow T^*\otimes Q_1$ is surjective with a kernel isomorphic to $H^2(g_1)$. This is a {\it crucial result} because it also proves that the additional CC has only to do with the the single second order component of the Riemann tensor in dimension $2$, a striking result that could not even be imagined by standard methods. \\
\end{proof}

Now, we have explained why the new zero order PD equation ${\xi}_1=0 \Leftrightarrow {\xi}^1=0$ should be replaced by the condition 
$x^i{\xi}_i=0$ by using the space euclidan metric for lowering the indices. Differentiating with respect to $x^j$, we obtain ${\delta}^i_j {\xi}_i + x^i {\xi}_{i,j}=0$ with the kronecker symbol $\delta$ and, contracting with $x^j$ we finally get $x^j{\xi}_j +x^ix^j{\xi}_{i,j}=0$. hence {\it on space}, we get a new subsystem by adding to the standard Killing system of space $(n=3)$ the above zero order constraint in order to get a system $R'_1$ with $dim(R'_1)= (3+9) - (6+2+1)=3$. Coming back to the computation previously done with the Schwarzschild metric while using only ${\xi}^0$ and ${\xi}^1$, we discover that {\it the new system does not any longer depend on} "$A$" (See other examples[25]).  \\

Its study can be therefore replaced by that of the $3$-dimensional system $R'_1$ which is defining a nontransitive system of infinitesimal lie equations, that is the map ${\pi}^1_0:R'_1 \rightarrow T$ is nonlonger surjective, a result modyfying the constructions of the Vessiot structure equations but this is out of our story. Supressing from now on the " ' " for simplicity, we are thus led to $n=3$ and the 
{\it  formally integrable} system of $9$ linearly independent equations equations were $\omega$ is the Euclidean metric: \\
\[   {\xi}_{i,j}+{\xi}_{j,i}=0,\hspace{1cm}   x^i{\xi}_{i,j } + {\xi}_j=0, \hspace{1cm} x^i{\xi}_i=0  \]
because $x^j(x^i {\xi}_{i,j} +{\xi}_j)= x^ix^j{\xi}_{i,j} +x^i{\xi}_i=x^i{\xi}_i=0$.  \\
Changing slighty the notations with now $n=3, m=dim(E)=2$ in order to keep an upper index for any unknown while setting ${\xi}^3=-\frac{x^1}{x^3}{\xi}^1 - \frac{x^2}{x^3}{\xi}^2$, we get the following system $R_1\subset J_1(E)$ with $dim(R_1)=3$ because $par_1=\{{\xi}^1,{\xi}^2, {\xi}^2_1\}$ and corresponding Janet tabular:  \\
\[ \left\{   \begin{array}{rcl}
{\Phi}^5 & \equiv & {\xi}^2_3 + \frac{x^1}{x^3} {\xi}^2_1 - \frac{1}{x^3}{\xi}^2 =0  \\
{\Phi}^4 & \equiv & {\xi}^1_3 - \frac{x^2}{x^3} {\xi}^2_1 - \frac{1}{x^3} {\xi}^1 =0  \\
{\Phi}^3 & \equiv & {\xi}^2_2=0  \\
{\Phi}^2 & \equiv & {\xi}^1_2 + {\xi}^2_1=0    \\
{\Phi}^1 & \equiv & {\xi}^1_1=0  
\end{array}  \right.  \fbox{  $  \begin{array} {ccc}
1 & 2 & 3  \\
1 & 2 & 3   \\
1 & 2 & \bullet  \\
1 & 2 & \bullet  \\
1 & \times & \bullet  
\end{array}  $  }  \] 
It is easy to check that all the second order jets vanish and that the general solution $\{ {\xi}^1= - ax^2 + cx^3, {\xi}^2= ax^1 - bx^3\}\Rightarrow {\xi}^3= - cx^1 + b x^2$ depends on $3$ arbitrary constants $(a,b,c)$ in such a way that the three space rotations are separately and respectively obtained by each element of the basis $\{  (1,0,0),(0,1,0),(0,0,1) \}$.\\

As before but with a different system, we have the following commutative diagrams:  

 \[  \begin{array}{rccccccccccl}
 &&&& 0 &&0&&0&  & \\
 &&&& \downarrow && \downarrow && \downarrow  & & \\
  && 0 & \longrightarrow &S_3T^*\otimes E&\longrightarrow&S_2T^*\otimes F_0 &\longrightarrow & h_2 & \longrightarrow  0  \\
   &&\downarrow && \downarrow & & \downarrow & & \downarrow & &    & \\
0& \longrightarrow &R_3&\longrightarrow &J_3(E)& \longrightarrow &J_2(F_0) &\longrightarrow & F_1 &   \longrightarrow 0 \\
    & & \downarrow && \hspace{5mm}\downarrow {\pi}^3_4 & &\hspace{5mm} \downarrow {\pi}^2_1 & & \downarrow  &  \\
   0 & \longrightarrow & R_2& \longrightarrow & J_2(E) & \longrightarrow &J_1(F_0)& \longrightarrow &Q_1& \longrightarrow  0 \\
   && \downarrow && \downarrow & & \downarrow & & \downarrow &      \\
   &&0&& 0 && 0 && 0  &  
   \end{array}     \]
 \[  \begin{array}{rccccccccccl}
 &&&& 0 &&0&& 0 &  & \\
 & &&& \downarrow && \downarrow &&\downarrow &  & \\
  && 0 & \longrightarrow &20 &\longrightarrow& 30 &\longrightarrow & 10 &\longrightarrow  0  \\
   &&\downarrow && \downarrow & & \downarrow & & \downarrow & &    & \\
0& \longrightarrow &3&\longrightarrow & 40 & \longrightarrow & 50 &\longrightarrow & 13 &   \longrightarrow 0 \\
    & & \downarrow && \hspace{5mm}\downarrow {\pi}^3_2 & &\hspace{5mm} \downarrow {\pi}^2_1 & & \downarrow  &  \\
   0 & \longrightarrow & 3 & \longrightarrow & 20 & \longrightarrow & 20 & \longrightarrow &  3  & \longrightarrow  0 \\
   &&\downarrow && \downarrow & & \downarrow & & \downarrow &      \\
   && 0 && 0 && 0 && 0  &  
   \end{array}     \]

 \[  \begin{array}{rccccccccccl}
 &&&& 0 &&0&&&  & \\
 &&&& \downarrow && \downarrow && & & \\
  && 0 & \rightarrow &S_3T^*\otimes E&\rightarrow&S_2T^*\otimes F_0 &\rightarrow & h_2 &
   \rightarrow  0  \\
   &&&& \hspace{3mm} \downarrow \delta & & \hspace{3mm} \downarrow \delta & & \downarrow & &    & \\
&  &0&\rightarrow &T^*\otimes S_2T^*\otimes E & \rightarrow &T^*\otimes T^* \otimes F_0 &\rightarrow & T^*\otimes h_1 &   \rightarrow 0 \\
    & & && \hspace{3mm}\downarrow \delta  & &\hspace{3mm} \downarrow \delta& & \downarrow  &  \\
   0 & \rightarrow & {\wedge}^2T^*\otimes g_1& \rightarrow & {\wedge}^2T^*\otimes T^* \otimes E & \rightarrow & {\wedge}^2T^*\otimes F_0  & \rightarrow & 0 &  \\
   && \hspace{3mm} \downarrow \delta && \hspace{3mm} \downarrow \delta & & \downarrow & & &   &   \\
 0   & \rightarrow & {\wedge}^3T^*\otimes E  & = & {\wedge}^3T^*\otimes E & \rightarrow & 0  & &&&   \\
       &   &  \downarrow &  & \downarrow  &  &  &  &  &  &  \\
   &&0&& 0 && && &  &
   \end{array}     \]

 \[  \begin{array}{rccccccccccl}
 &&&& 0 &&0&&&  & \\
 &&&& \downarrow && \downarrow && & & \\
  && 0 & \rightarrow &20&\rightarrow&30 &\rightarrow & 10 &
   \rightarrow  0  \\
   &&&& \hspace{3mm} \downarrow \delta & & \hspace{3mm} \downarrow \delta & & \downarrow & &    & \\
&  &0&\rightarrow &36 & \rightarrow &45 &\rightarrow & 9 &   \rightarrow 0 \\
    & & && \hspace{3mm}\downarrow \delta  & &\hspace{3mm} \downarrow \delta& & \downarrow  &  \\
   0 & \rightarrow & 3 & \rightarrow & 18 & \rightarrow & 15 & \rightarrow & 0 &  \\
   && \hspace{3mm} \downarrow \delta && \hspace{3mm} \downarrow \delta & & \downarrow & & &   &   \\
 0   & \rightarrow &  2  & = &  2  & \rightarrow & 0  & &&&   \\
       &   &  \downarrow &  & \downarrow  &  &  &  &  &  &  \\
   &&0&& 0 && && &  &
   \end{array}     \]

The next result points out the importance of the Spencer $\delta$-cohomology:  \\

\noindent 
{\bf LEMMA 3B.4}: The last symbol diagram is commutative and exact. In particular, the lower left map $\delta$ is surjective and thus  the upper right induced map $h_2 \rightarrow T^*\otimes Q_1$ is  also surjective while these two maps have isomorphic kernels.  \\

\begin{proof}
The $3$ components of ${\wedge}^2T^*\otimes g_1$ are $\{ v^2_{1, 12}, v^2_{1,13}, v^2_{1,23}\}$ and the map $\delta$ is described by the two linear equations:  \\
\[  v^1_{1,23} +v^1_{2,31} + v^1_{3,12}=0, \hspace{5mm}  v^2_{1,23} +v^2_{2,31} + v^2_{3,12}=0 \]
that is to say by the two linearly independent equations:  \\
\[    v^2_{1,13} + \frac{x^2}{x^3}v^2_{1,12}=0, \hspace{5mm} v^2_{1,23} - \frac{x^1}{x^3}v^2_{1,12}=0 \]
Accordingly, in the left column we have:  \\
\[    dim(H^2(g_1))=dim(Z^2(g_1))=dim(ker(\delta))=1\]   
An unusual snake-type diagonal chase left to the reader as an exercise proves that the induced map $h_2\rightarrow T^*\otimes Q_1$ is surjective with a kernel isomorphic to $H^2(g_1)$. This is indeed a {\it crucial result} because it also proves that the additional CC has only to do with the the single second order component of the Riemann tensor in dimension $2$, a striking result that could not even be imagined by standard methods. \\
\end{proof}   
We notice that it is a pure chance that, {\it in this example}, we could have set $F_1=Q_2$. Indeed, considering $B_r=ker(J_r(F_0)\rightarrow Q_r)$, we have $dim(B_1)= 20-3=1$ and $dim(B_2)=40-3=50-13=37$. However, we have ${\rho}_1(B_1)=ker (J_2(F_0)\rightarrow J_1(Q_1))$ and, {\it as this map is known to be surjective}, we have $dim({\rho}_1(B_1))= 50 - 12=38$ and thus the strict inclusion $B_2\subset {\rho}_1(B_1)$ proving the existence of a new third order CC that is {\it not} a differential consequence of the known differentially independent second order CC. The reader may compare this situation to the one met in the $4$-dimensional framework and understand why it is not possible to avoid the tedious direct checking we had to provide.  \\

Collecting the above results, we find the $3$ {\it first order} differentially independent generating CC coming from the Janet tabular and the {\it additional} single {\it second order} generating CC describing the $2$-dimensional {\it Riemann operator}, that is the linearized Riemann tensor in the space $(x^1,x^2)$:  \\

\[  \left\{     \begin{array}{rcl}
{\Psi}^4 & \equiv & d_{22}{\Phi}^1 + d_{11}{\Phi}^3 - d_{12}{\Phi}^2  =0  \\
              &           &                        \\
{\Psi}^3 & \equiv &   d_3{\Phi}^3 - d_2{\Phi}^5 + \frac{x^1}{x^3}d_1 {\Phi}^3 - \frac{1}{x^3}{\Phi}^3=0    \\
{\Psi}^2 & \equiv &   d_3{\Phi}^2 - d_2{\Phi}^4 - d_1{\Phi}^5  - \frac{x^2}{x^3} d_1{\Phi}^3 + \frac{x^1}{x^3} (d_1{\Phi}^2 - d_2{\Phi}^1) - \frac{1}{x^3} {\Phi}^2 =0        \\
{\Psi}^1 & \equiv & d_3{\Phi}^1 - d_1{\Phi}^4 - \frac{x^2}{x^3} (d_1{\Phi}^2 - d_2{\Phi}^1) - \frac{1}{x^3} {\Phi}^1  =0
\end{array} \right.  \] 

An elementary but tedious computation provides the second order CC:  \\
\[   x^3 (d_{22}{\Psi}^1 +d_{11}{\Psi}^3 - d_{12}{\Psi}^2) - (x^1d_1{\Psi}^4 + x^2d_2{\Psi}^4 + x^3d_3{\Psi}^4) -  2 {\Psi}^4 =0   \]

The corresponding differential sequence written with differential modules over the ring $D=K[d_1,d_2,d_3]$ with $K=\mathbb{Q}(x^1,x^2,x^3)$ is:  \\
\[          0 \rightarrow D  \underset{2}{\rightarrow} D^4 \underset{2}{\rightarrow} D^5 \underset{1}{\rightarrow} D^2  \stackrel{p}{\rightarrow}M 
\rightarrow 0   \]
where $p$ is the canonical (residual) projection. We check indeed that $ 1 - 4 + 5 -2 = 0$ but this sequence is quite far from being even strictly exact. Of course, as $R_2$ is involutive, we may set 
$C_r={\wedge}^rT^*\otimes R_2$ and obtain the corresponding canonical second Spencer sequence which is induced by the Spencer operator:\\ 
\[   0  \rightarrow \Theta \stackrel{j_2}{\longrightarrow} C_0 \stackrel{D_1}{\longrightarrow} C_1 \stackrel{D_2}{\longrightarrow}  C_2 \stackrel{D_3}{\longrightarrow}C_3 \rightarrow 0    \]  with dimensions:  \\
\[   0  \rightarrow \Theta \stackrel{j_2}{\longrightarrow} 3 \underset{1}{\stackrel{D_1}{\longrightarrow}} 9 \underset{1}{\stackrel{D_2}{\longrightarrow}}  9 \underset{1}{ \stackrel{D_3}{\longrightarrow}} 3 \rightarrow 0    \]
Proceeding inductively as we did for finding the second order CC, we may obtain by combinatorics  the following formally exact sequence:  \\
\[  0 \rightarrow  \Theta \rightarrow 2 \underset 1{\rightarrow} 5 \underset 2{\rightarrow}13 \underset 1{\rightarrow} 19 \underset 1{\rightarrow} 12 \underset 1{\rightarrow} 3 \rightarrow 0  \]
with Euler-Poincar\'{e} characteristic $2 - 5 + 13 - 19 + 12 - 3=0$ but, as before, there is a matrix $260 \times 280$ at least and we doubt about the use of computer algebra, even on such an elementary example. We also let the reader compute the corresponding Janet sequence as a first step towards the {\it Vessiot structure equations} which are not easily obtained because we have now: \\ 
\[  x^i {\xi}_i=x^i {\omega}_{ij} {\xi}^j \Rightarrow {\xi}_{i,j} + {\xi}_{j,i} - 2 {\gamma}^r_{ij}{\xi}_r=0\]
where $\gamma$ denotes the Christoffel symbols.  \\
This result justifies " {\it a fortiori} " the comments we have already provided. The reader may compare such an example with the Janet example (where we have one third order CC and one sixth order additional CC) with the major difference that we have now a formally integrable system. We do not know any other similar situation. \\

In order to achieve the study of the Schwarzschild metric in a purely intrinsic way, we prove that the situation of the previous example is just decribing the way to exhibit the torsion part $t(M)\subseteq M$ of a differential module by computing a certain extension module. Coming back to the systems already obtained and keeping in mind that ${\xi}_1=0 \Rightarrow {\xi}_{0,0}=0, {\xi}_{0,1}- \frac{A'}{A}{\xi}_0=0, {\xi}_{0,2}=0, {\xi}_{0,3}=0  \Rightarrow {\xi}^0=cst$ while replacing $r^2{\xi}_2, r^2{\xi}_3$ by ${\xi}^2, {\xi}^3$ respectively, we may therefore replace the integration of the previous system by that of the simpler system:  \\ 
\[ \left\{   \begin{array}{lcl}
{\Phi}^5 & \equiv & {\xi}^3_1 =0  \\
{\Phi}^4 & \equiv & {\xi}^2_1 =0 \\
{\Phi}^3 & \equiv & {\xi}^3_3+ sin(\theta)cos(\theta) {\xi}^2 =0 \\
{\Phi}^2 & \equiv &  {\xi}^2_3 + {\xi}^3_2 - 2cot(\theta)\, {\xi}^3  =0  \\
{\Phi}^1 & \equiv & {\xi}^2_2 =0 
\end{array}  \right.    \fbox{  $  \begin{array} {ccc}
2 & 3 & 1  \\
2 & 3 & 1 \\
2 & 3 & \bullet   \\
2 & 3 & \bullet  \\
2 & \times  & \bullet
\end{array}  $  }  \] 
allowing to define an isomorphic differential module because both systems are formally integrable though not involutive, with the same dimension $2+(2\times 3) -5=3$ with ${par}_2={par}_1=\{ {\xi}^2,{\xi}^3, {\xi}^3_2\}$. We have now similarly the $3$ {\it first order} CC and the single {\it second order} CC:  \\
\[  \left\{     \begin{array}{rcl}
{\Psi}^4 & \equiv &    d_{33}{\Phi}^1 +d_{22}{\Phi}^3 - d_{23}{\Phi}^2\\
              &             & - sin(\theta)cos(\theta) d_2{\Phi}^1 - 2 cot(\theta) d_2{\Phi}^3 + 2 sin^2(\theta) {\Phi}^1 + \frac{2}{sin^2(\theta)} {\Phi}^3 =0   \\                        \\
{\Psi}^3 & \equiv &  d_1{\Phi}^3 - d_3{\Phi}^5 - sin(\theta)cos(\theta) {\Phi}^4 =0  \\
{\Psi}^2 & \equiv &  d_1{\Phi}^2 - d_3{\Phi}^4 -d_2{\Phi}^5 - 2cot(\theta){\Phi}^5 =0   \\
{\Psi}^1 & \equiv & d_1{\Phi}^1 - d_2{\Phi}^4=0
\end{array} \right.  \] 
describing again the single component of the linearized Riemann tensor for $(\theta,\phi)$ and the first Spencer cohomology group $H^2(g_1)$ of the first symbol $g_1\subset T^*\otimes E$ with $dim(E)=2$ and $dim(H^2(g_1))=dim ({\wedge}^2T^*\otimes g_1)- dim({\wedge}^3T^*\otimes E)= 3 - 2=1 $. Of course, we could even simplified the later system by considering the new system:  \\
\[ \left\{   \begin{array}{lcl}
{\Phi}^3 & \equiv & {\xi}^3_3+ sin(\theta)cos(\theta) {\xi}^2 =0 \\
{\Phi}^2 & \equiv &  {\xi}^2_3 + {\xi}^3_2 - 2cot(\theta)\, {\xi}^3  =0  \\
{\Phi}^1 & \equiv & {\xi}^2_2 =0 
\end{array}  \right.    \fbox{  $  \begin{array} {cc}
2 & 3   \\
2 & 3 \\
2 &  \bullet   
\end{array}  $  }  \] 
We let the reader fill in the details and discover again the only CC ${\Psi}^4=\Psi=0$.    \\
Considering both situations already studied with $n=3,m=2$, we discover that the differential modules defined by the system ${\Psi}^1=0,{\Psi}^2=0, {\Psi}^3=0$ are isomoprphic, provided we extend conveniently the ground differential field. Hence, in both cases, we have $t(M)\neq 0$ and the torsion submodule has a {\it single generator} namely the residue of ${\Psi}^4$ which is satisfying for example the so-called {\it autonomous} equation:\\
\[   x^1 d_1{\Psi}^4 + x^2 d_2 {\Psi}^4 + x^3 d_3 {\Psi}^4 + 2 {\Psi}^4=0   \]
Setting $M'=M/t(M)$, we have the short exact sequence:  \\
\[     0 \rightarrow t(M) \rightarrow M  \rightarrow M' \rightarrow 0  \]
where the torsion-free differential module $M'$ is now defined by ${\Psi}^1=0, ..., {\Psi}^4=0$. The explanation can only be natural in the framework of differential homological algebra because, {\it by construction}, the CC operator ${\cal{D}}_1$ is parametrized by ${\cal{D}}$ and MUST therefore provide the torsion-free differential module $M'$. This result explains for the first time the intrinsic character of the additional higher order generating CC that can be found in an apparently strange manner. In a more detailed way, let us proceed as follows in order to construct the following commutative diagram:   \:
\[  \begin{array}{rcccl}
  & & & &   4  \\
  & & & \stackrel{{\cal{D}}'_1}{\nearrow} &    \\
 2 & \stackrel{{\cal{D}}}{\longrightarrow} &  5  & \stackrel{{\cal{D}}_1}{\longrightarrow} & 3      \\
   &  &  &  &    \\
   &  &  &  &    \\
  2 & \stackrel{ad({\cal{D}})}{\longleftarrow} & 5 & \stackrel{ad({\cal{D}}_1)}{\longleftarrow} &3
  \end{array}  \]
\noindent
1) Write down the operator ${\cal{D}}_1:({\Phi}^1, ..., {\Phi}^5) \rightarrow ({\Psi}^1,{\Psi}^2, {\Psi}^3)$ as we did.  \\
\noindent
2) Multiply by test functions $\lambda= ({\lambda}^1,{\lambda}^2,{\lambda}^3)$, integrate by parts and construct the formal adjoint $ad({\cal{D}}_1):({\lambda}^1,{\lambda}^2,{\lambda}^3) \rightarrow ({\mu}^1, ..., {\mu}^5)$:   \\
\[ \left\{ \begin{array}{lcl}
-d_1{\lambda}^1 & = & {\mu}^1   \\
-d_1{\lambda}^2 & = & {\mu}^2    \\
-d_1{\lambda}^3 & = & {\mu}^3   \\
d_3{\lambda}^2+ d_2{\lambda}^1 - sin(\theta)cos(\theta){\lambda}^3 & = & {\mu}^4  \\
d_3{\lambda}^3+ d_2{\lambda}^2 - 2 cos(\theta){\lambda}^2 & = & {\mu}^5
\end{array} \right.  \]
\noindent
3) Construct generating CC as an operator $ad({\cal{D}}): ({\mu}^1,...,{\mu}^5) \rightarrow ({\nu}^1,{\nu}^2) $:   \\
\[ \left\{ \begin{array}{lcl}
d_1{\mu}^4 + d_3{\mu}^2 +d_2{\mu}^1 - sin(\theta)cos(\theta){\mu}^3 & = &  {\nu}^1  \\
d_1{\mu}^5 + d_3 {\mu}^3 + d_2 {\mu}^2  - 2 cos(\theta){\mu}^2 & = & {\nu}^2
\end{array} \right. \]
\noindent
4) Exhibit $ad(ad({\cal{D}}))={\cal{D}}: ({\Upsilon}^1, {\Upsilon}^2) \rightarrow ({\Phi}^1,...,{\Phi}^5)$:  \\
\[  \left\{  \begin{array}{lcl}
-d_2 {\Upsilon}^1 & = & {\Phi}^1    \\
-d_3{\Upsilon}^1 -d_2{\Upsilon}^2 - 2 cos(\theta) {\Upsilon}^2 & = & {\Phi}^2    \\
-d_3 {\Upsilon}^2- sin(\theta)cos(\theta) {\Upsilon}^1 & = & {\Phi}^3  \\
-d_1 {\Upsilon}^1 & = & {\Phi}^4    \\
-d_1 {\Upsilon}^2 & = & {\Phi}^5
\end{array}  \right.  \]
\noindent
5) Construct generating CC $ {\cal{D}}':({\Phi}^1,..,{\Phi}^5)$ and compare ${\cal{D}} \leq {\cal{D}}'$. Then each new CC is a torsion element of the differential module determined by ${\cal{D}}$ which is thus parametrized by ${\cal{D}}_{-1}$ if and only if ${\cal{D}}'={\cal{D}}$. In the present case, ${\Psi}^1,{\Psi}^2,{\Psi}^3$ being differentially independent, we find the only additional generating CC ${\Psi}^4=0$. \\
Accordingly, the situation in GR cannot evolve as long as people will not acknowledge the fact that the components of the Weyl tensor are {\it similarly} playing the part of torsion elements (the so-called {\it Lichnerowicz waves} in [6,12]) for the equations $Ricci=0$, a result only depending on the group structure of the conformal group of space-time that brings the splitting $Riemann = Weyl \oplus Ricci $.   \\
 
\vspace{3cm}

\noindent
{\bf C) KERR METRIC}:\\

We now write the Kerr metric in Boyer-Lindquist coordinates:  \\
\[  \begin{array}{rcl}
ds^2  &  =  & \frac{{\rho}^2 - mr}{{\rho}^2}dt^2 - \frac{{\rho}^2}{\Delta} dr^2  -  {\rho}^2 d{\theta}^2  \\
  &    &  + \frac{2a m r sin^2(\theta)}{{\rho}^2} dtd\phi -  \frac{(r^2+a^2)^2 - a^2 \Delta sin^2(\theta)}{{\rho}^2}sin^2(\theta) d{\phi}^2          
  \end{array}   \]
where we have set $  \Delta= r^2  -mr +a^2 , \,\,  {\rho}^2=r^2 + a^2 cos^2(\theta) $ and we check that:  \\
\[  a=0  \Rightarrow  ds^2= (1- \frac{m}{r})dt^2 - \frac{1}{1- \frac{m}{r}}dr^2 -r^2d{\theta}^2 - r^2 sin(\theta)^2 d{\phi}^2 \]
as a well known way to recover the Schwarschild metric. Now, we notice that $t$ or $\phi$ do not appear in the coefficients of the metric and thus, as the maximum subgroup of invariance of the Kerr metric {\it must} be contained in the maximum subgroup of invariance of the Schwarzschild metric because of the above limit when $a\rightarrow 0$, we obtain the only possible $2$ infinitesimal generators $\{ {\partial}_t, {\partial}_{\phi}\}$ and we have the fundamental diagram I with fiber dimensions:  \\
\[   \begin{array}{rcccccccccccccl}
 &&&&& 0 &&0&&0&& 0 &  &0&  \\
 &&&&& \downarrow && \downarrow && \downarrow & & \downarrow &   & \downarrow &  \\
  & 0& \rightarrow& \Theta &\stackrel{j_2}{\rightarrow}&2 &\stackrel{D_1}{\rightarrow}& 8 &\stackrel{D_2}{\rightarrow} & 12 &\stackrel{D_3}{\rightarrow}& 8 &\stackrel{D_4}{\rightarrow}& 2 &\rightarrow 0 \\
  &&&&& \downarrow & & \downarrow & & \downarrow & & \downarrow &&\downarrow &     \\
   & 0 & \rightarrow & 4 & \stackrel{j_2}{\rightarrow} & 60 & \stackrel{D_1}{\rightarrow} & 160 &\stackrel{D_2}{\rightarrow} & 180 &\stackrel{D_3}{\rightarrow} & 96 &\stackrel{D_4}{\rightarrow} &20&   \rightarrow 0 \\
   & & & \parallel && \downarrow  & & \downarrow  & & \downarrow  & & \downarrow & & \downarrow  & \\
   0 \rightarrow & \Theta &\rightarrow & 4 & \stackrel{\cal{D}}{\rightarrow} & 58 & \stackrel{{\cal{D}}_1}{\rightarrow} & 152 & \stackrel{{\cal{D}}_2}{\rightarrow} & 168 & \stackrel{{\cal{D}}_3}{\rightarrow} & 88 &\stackrel{{\cal{D}}_4}{\rightarrow} & 18 & \rightarrow  0 \\
   &&&&& \downarrow & & \downarrow & & \downarrow & & \downarrow &  &\downarrow &   \\
   &&&&& 0 && 0 && 0 &&0 &&0 &  
   \end{array}     \]
with Euler-Poincar\'{e} characteristic $4 - 58 +152 - 168 + 88 - 18 = 0  $. Comparing the {\it surprisingly high} dimensions of the Janet bundles with the {\it surprisingly low} dimensions of the Spencer bundles needs no comment on the physical usefulness of the Janet sequence, despite its purely mathematical importance. In addition, using now the same notations as in the preceding section, we have the additional {\it zero order} equations 
${\xi}^1=0, {\xi}^2=0$ produced by the non-zero components of the Weyl tensor and thus, {\it at best}, $dim(R^{(3)}_0)=2 \Leftrightarrow  dim((R^{(2)}_1)=2$, if the zero order equations are obtained after {\it only} two prolongations. As these zero order equations depend on $j_2(\Omega)$, 
{\it at best}, we should obtain therefore eventually $dim(Q_2)=10+2=12$ CC of order $2$ and $60-(4\times 12)=12$ CC of order $3$ at least.  \\ 

Using finally cartesian coordinates with ${\xi}^3=0, \,\,x^1 {\xi}^1 +x^2{\xi}^2=0$, we have only to study the following first order involutive system for ${\xi}^1=\xi$ with coefficients no longer depending on $(a,m)$, providing the only generator $x^1{\partial}_2 - x^2 {\partial}_1$: \\
\[ \left\{   \begin{array}{lcl}
{\Phi}^3 & \equiv & {\xi}_3 =0 \\
{\Phi}^2 & \equiv &  {\xi}_2 - \frac{1}{x^2}{\xi}  =0  \\
{\Phi}^1 & \equiv & {\xi}_1 =0 
\end{array}  \right.    \fbox{  $  \begin{array} {ccc}
1 & 2 & 3   \\
1 & 2 & \bullet \\
1 & \bullet &  \bullet   
\end{array}  $  }  \] 

\[  {\Psi}^3 \equiv d_3{\Phi}^2- d_2{\Phi}^3 + \frac{1}{x^2}{\Phi}^3=0, {\Psi}^2\equiv d_3{\Phi}^1 -d_1{\Phi}^3=0,
           {\Psi}^1\equiv d_2{\Phi}^1-d_1{\Phi}^2 - \frac{1}{x^2}{\Phi}^1=0  \]
\[   \Rightarrow   \hspace{1cm}   d_3{\Psi}^1- d_2{\Psi}^2+d_1{\Psi}^3+\frac{1}{x^2}{\Psi}^2 =0  \]

\[   \begin{array}{rcccccccccccl}
 &&&&& 0 &&0&&0&  &0&  \\
 &&&&& \downarrow && \downarrow && \downarrow &    & \downarrow &  \\
  & 0& \rightarrow& \Theta &\stackrel{j_1}{\rightarrow}&1 &\stackrel{D_1}{\rightarrow}& 3 &\stackrel{D_2}{\rightarrow} & 3 &\stackrel{D_3}{\rightarrow}& 1 &\rightarrow 0 \\
  &&&&& \downarrow & & \downarrow & & \downarrow & &\parallel &     \\
   & 0 & \rightarrow & 1 & \stackrel{j_1}{\rightarrow} & 4 & \stackrel{D_1}{\rightarrow} & 6 &\stackrel{D_2}{\rightarrow} & 4 &\stackrel{D_3}{\rightarrow} & 1 &   \rightarrow 0 \\
   & & & \parallel && \downarrow  & & \downarrow  & & \downarrow  &  & \downarrow  & \\
   0 \rightarrow & \Theta &\rightarrow & 1 & \stackrel{\cal{D}}{\rightarrow} & 3 & \stackrel{{\cal{D}}_1}{\rightarrow} & 3 & \stackrel{{\cal{D}}_2}{\rightarrow} & 1 & \rightarrow & 0 & \\
   &&&&& \downarrow & & \downarrow & & \downarrow &  &   &   \\
   &&&&& 0 && 0 && 0 && &  
   \end{array}     \]

This final result definitively proves that, {\it as far as differential sequences are concerned}, the only important object is the group, not the metric.  \\

\newpage
\noindent
{\bf 4) CONCLUSION}  \\

We may summarize the results obtained in the $3$ previous subsections by saying:  \\
 \[    JANET \,\,\,AND \,\,\,SPENCER \,\,\,PLAY \,\,\,AT \,\,\,SEE-SAW   \]
because we have the formula $dim(C_r) + dim(F_r) = dim(C_r(E)) $ and the sum thus only depends on $(n,m,q)$ with $n=dim(X), m=dim(E)$ and $q$ is the order of the involutive operator allowing to construct the sequences, but not on the underlying Lie group or Lie pseudogroup group when $E=T$. Hence, the smaller is the background group, the smaller are the dimensions of the Spencer bundles and the higher are the dimensions of the Janet bundles. As a byproduct, we claim that the only solution for escaping is to increase the dimension of the Lie group involved, adding successively $1$ dilatation and $4$ elations in order to deal with the conformal group of space-time while using the Spencer sequence instead of the Janet sequence ([30],[42]). Finally, taking into account the partition $10=4+4+2$ into three blocks of the Ricci tensor that we obtained, it follows that Einstein equations are not mathematically coherent with group theory and  formal integrability. Two forthcoming publications will achieve this game.  \\

\vspace{3cm}

\noindent
{\bf REFERENCES}  \\

\noindent
[1] Aksteiner, S., B\"{a}ckdahl, T.: All Local Gauge Invariants for Perturbations of the Kerr Spacetime, 2018, arxiv:1803.05341.   \\
\noindent
[2] Andersson, L.: Spin Geometry and Conservation Laws in the Kerr Spacetime, 2015, arxiv:1504.02069.  \\
\noindent
[3] Assem, I.: Alg\'{e}bres et Modules, Masson, Paris, 1997.  \\
\noindent
[4] Bjork, J.E.: Analytic D-Modules and Applications, Kluwer, 1993.  \\ 
\noindent
[5] Bourbaki, N.: Alg\'{e}bre, Ch. 10, Alg\'{e}bre Homologique, Masson, Paris, 1980.\\
\noindent
[6] Choquet-Bruhat, Y.: (2015) Introduction to General Relativity, Black Holes and Cosmology, Oxford University Press.  \\
\noindent
[7] Eisenhart, L.P.: Riemannian Geometry, Princeton University Press, 1926.  \\
\noindent
[8] Foster, J., Nightingale, J.D. (1979) A Short Course in General relativity, Longman.  \\
\noindent
[9] Goldschmidt, H.: Prolongations of Linear Partial Differential Equations: I A Conjecture of Elie Cartan, Ann. Scient. Ec. Norm. Sup., 4, 1 (1968) 417-444.  \\
\noindent
[10] Goldschmidt, H.: Prolongations of Linear Partial Differential Equations: I Inhomogeneous equations, Ann. Scient. Ec. Norm. Sup., 4, 1 (1968) 617-625.  \\
\noindent
[11] Hu,S.-T.: Introduction to Homological Algebra, Holden-Day, 1968.  \\
\noindent
[12] Hughston, L.P., Tod, K.P. (1990) An Introduction to General Relativity, London Math. Soc. Students Texts 5, Cambridge University Press . \\
\noindent
[13] Janet, M.: Sur les Syst\`{e}mes aux D\'{e}riv\'{e}es Partielles, Journal de Math., 8(3) (1920) 65-151.  \\
\noindent
[14] Kashiwara, M.: Algebraic Study of Systems of Partial Differential Equations, M\'emoires de la Soci\'{e}t\'{e} Math\'ematique de France 63, 1995, 
(Transl. from Japanese of his 1970 Master's Thesis).\\
\noindent
[15] Khavkine, I.: The Calabi Complex and Killing Sheaf Cohomology, J. Geom. Phys., 113 (2017) 131-169. \\
arxiv:1409.7212  \\
\noindent
[16] Khavkine, I.: Compatibility Complexes of Overdetermined PDEs of Finite Type, with Applications to to the Killing Equation, 2018. \\
arxiv:1805.03751  \\
\noindent
[17] Kunz, E.: Introduction to Commutative Algebra and Algebraic Geometry, Birkh\"{a}user, Boston, 1985.  \\
\noindent
[18] Macaulay, F.S.: The Algebraic Theory of Modular Systems, Cambridge, 1916.  \\
\noindent  
[19] Northcott, D.G.: An Introduction to Homological Algebra, Cambridge university Press, 1966.  \\
\noindent
[20] Pommaret, J.-F.: Systems of Partial Differential Equations and Lie Pseudogroups, Gordon and Breach, New York; Russian translation: MIR, Moscow, 1978.\\
\noindent
[21] Pommaret, J.-F.: Differential Galois Theory, Gordon and Breach, New York, 1983.\\
\noindent
[22] Pommaret, J.-F.: Lie Pseudogroups and Mechanics, Gordon and Breach, New York, 1988.\\
\noindent
[23] Pommaret, J.-F.: Partial Differential Equations and Group Theory, Kluwer, 1994.\\
http://dx.doi.org/10.1007/978-94-017-2539-2    \\
\noindent
[24] Pommaret, J.-F.: Dualit\'{e} Diff\'{e}rentielle et Applications, C. R. Acad. Sc. Paris, 320, S\'{e}rie I (1995) 1225-1230.  \\
\noindent
[25] Pommaret, J.-F.: Partial Differential Control Theory, Kluwer, Dordrecht, 2001 (1000 pp).\\
\noindent  
[26] Pommaret, J.-F.: Relative Parametrization of Linear Multidimensional Systems, Multidim. Syst. Sign. Process. (MSSP), Springer, 26 
(2013) 405-437. \\
http://dx.doi.org/10.1007/s11045-013-0265-0  \\
\noindent
[27] Pommaret, J.-F.: The Mathematical Foundations of General Relativity Revisited, Journal of Modern Physics, 4 (2013) 223-239.\\
http://dx.doi.org/10.4236/jmp.2013.48A022   \\
\noindent
[28] Pommaret, J.-F.: The Mathematical Foundations of Gauge Theory Revisited, Journal of Modern Physics, 5 (2014) 157-170.  \\
http://dx.doi.org/10.4236/jmp.2014.55026    \\
\noindent
[29] Pommaret, J.-F.: Deformation Theory of Algebraic and Geometric Structures, Lambert Academic Publisher, (LAP), 
Saarbrucken, Germany, 2016.  \\
http://arxiv.org/abs/1207.1964  \\
\noindent
[30] Pommaret, J.-F.: Why Gravitational Waves Cannot Exist, Journal of Modern Physics, 8,13 (2017) 2122-2158.  \\
http://dx.doi.org/10.4236/jmp.2017.813130   \\
\noindent
[31] Pommaret, J.-F.: Algebraic Analysis and Mathematical physics, 2017.  \\
http://arxiv.org/abs/1706.04105  \\
\noindent
[32] Pommaret, J.-F.: From Elasticity to Electromagnetism: Beyond the Mirror, \\
http://arxiv.org/abs/1802.02430  \\
\noindent
[33] Pommaret, J.-F.: New Mathematical Methods for Physics, NOVA Science Publisher, New York, 2018.  \\
\noindent  
[34] Quadrat, A. (2010) An Introduction to Constructive Algebraic Analysis and its Applications, 
Les cours du CIRM, Journees Nationales de Calcul Formel, 1(2), 281-471.\\
\noindent
[35] Quadrat, A., Robertz, R. (2014) A Constructive Study of the Module Structure of Rings of Partial Differential Operators, Acta Applicandae Mathematicae, 133, 187-234 (2014) 187-234. \\
http://hal-supelec.archives-ouvertes.fr/hal-00925533   \\
\noindent
[36] Rotman, J.J.: An Introduction to Homological Algebra, Pure and Applied Mathematics, Academic Press, 1979.  \\
\noindent
[37] Schneiders, J.-P.: ( 1994): An Introduction to D-Modules, Bull. Soc. Roy. Sci. Li\'{e}ge, 63, 223-295.  \\
\noindent
[38] Shah, A.G., Whitting, B.F., Aksteiner, S., Andersson, L., B\"{a}ckdahl, T.: Gauge Invariant perturbation of Schwarzschild Spacetime, 2016, 
arxiv:1611.08291  \\
\noindent
[39] Spencer, D.C.: Overdetermined Systems of Partial Differential Equations, Bull. Amer. Math. Soc., 75 (1965) 1-114.\\
\noindent
[40] Vessiot, E.: Sur la Th\'{e}orie des Groupes Infinis, Ann. Ec. Normale Sup., 20 (1903) 411-451 (Can be obtained from  http://numdam.org). \\
\noindent
[41] Vessiot, E. (1904) Sur la Th\'{e}orie de Galois et ses Diverses G\'{e}n\'{e}ralisations, Ann. Ec. Norm. Sup., 21 (1904) 9-85 (Can be obtained from 
http://numdam.org).  \\
\noindent
[42] Weyl, H. (1918) Space, Time, Matter, Springer, 1958; Dover, 1952. \\
\noindent
[43] Zerz, E.: Topics in Multidimensional Linear Systems Theory, Lecture Notes in Control and Information Sciences (LNCIS) 256, 
Springer, 2000. \\

\end{document}